 \documentclass[final,5p,times,twocolumn]{elsarticle}
 
\usepackage{amssymb}
\usepackage[svgnames]{xcolor}
\usepackage{amsmath}
\usepackage{multirow}
\usepackage{placeins}
\usepackage[bookmarks=false]{hyperref}
\hypersetup{colorlinks,linkcolor={blue},citecolor={blue},urlcolor={red}}
\graphicspath{{Figures/}}

\begin{document}

\begin{frontmatter}



\title{Effect of polydispersity on the transport and sound absorbing properties of three-dimensional random fibrous structures}


\author[inst1,inst2,inst3,inst4]{Quang Vu Tran}

\affiliation[inst1]{organization={Univ Gustave Eiffel, Univ Paris Est Creteil},
            addressline={CNRS, UMR 8208, MSME}, 
            postcode={F-77454}, 
            state={Marne-la-Vallée},
            country={France}}

\author[inst1,inst4]{Camille Perrot}
\author[inst2,inst4]{Raymond Panneton}
\author[inst3]{Minh Tan Hoang}
\author[inst3]{Ludovic Dejaeger}
\author[inst3]{Valérie Marcel}
\author[inst3]{Mathieu Jouve}

\affiliation[inst2]{organization={Département de Génie Mécanique},
            addressline={Université de Sherbrooke}, 
            postcode={J1K 2R1}, 
            state={Québec},
            country={Canada}}
\affiliation[inst3]{organization={Adler Pelzer Group, Acoustic TechCenter R$\&$D},
            addressline={ Z.I. François Sommer – BP13}, 
            postcode={08210}, 
            state={Mouzon},
            country={France}}
\affiliation[inst4]{email={quang.vu.tran@usherbrooke.ca, camille.perrot@univ-eiffel.fr, raymond.panneton@usherbrooke.ca}}

\begin{abstract}
A technique is proposed that uses a multi-scale approach to calculate transport properties of compressed felts using only image analysis and numerical calculations. From the image analysis fiber diameter distribution and fiber orientation are determined. From a known porosity and the latter two characteristics, two representative elementary volumes (REV) are constructed: one based on the volume-weighted average diameter and one on an inverse volume-weighted average diameter. Numerical calculations on the former showed that it correctly estimates viscous and thermal permeabilities, while the latter correctly estimates tortuosity and viscous and thermal characteristic lengths. From these calculations, micro-macro analytical expressions are developed to estimate the transport properties of polydisperse composite felts based solely on open porosity, fiber diameter polydispersity, and fiber orientation. Good agreements are obtained between analytical predictions and measurements of transport properties.  The predicted transport properties are also used in the Johnson-Champoux-Allard-Lafarge (JCAL) equivalent fluid model to predict the sound absorption coefficient of the felts. Excellent agreements are obtained with impedance tube measurements. 
\end{abstract}




\begin{keyword}
Multiscale model\sep fibrous media \sep representative elementary volume \sep transport properties \sep sound absorption coefficient \sep compression effect\\
\vspace{0.25cm}

\textit{Credit:} This content may be downloaded for personal use only. Any other use requires prior permission of the author and the publisher. This material originally appeared in Q. V. Tran et al., Int. J. Solids Struct. 296, 112840 (2024) and may be found at \href{https://doi.org/10.1016/j.ijsolstr.2024.112840}{https://doi.org/10.1016/j.ijsolstr.2024.112840}
\end{keyword}

\end{frontmatter}


\section{Introduction}
\label{sec:Introduction}
Nonwoven fabrics are some of the most widespread man-made porous materials that are used in many engineering fields including health and medical care, energy or sound proofing applications. The main constituents of nonwoven are fibers that are linked together by cohesive bounds induced by the manufacturing process in the form of fibrous networks with transverse isotropy. Nonwoven fibrous media with a wide diversity of physical and mechanical properties (Dirrenberger \textit{et al.} \cite{DIRRENBERGER2014359},  Altendorf \textit{et al.} \cite{altendorf2014influence}, Bosco \textit{et al.} \cite{BOSCO201866}) can be manufactured by tailoring the nature of the raw materials and the manufacturing process conditions (e.g., type of geometry, bale opening and weighting of the fibers, fibers web creation, thermal bounding thickness adjustment and cutting). However, the links between composite nonwoven manufacturing parameters, the resulting fibrous microstructures, and the product performance are still not fully evidenced. For example, the permeability $k_0$ (Darcy \cite{darcy1856fontaines}) and the viscous characteristic length $\Lambda$ (Johnson \textit{et al.} \cite{johnson1986new}) of  felts often follow a nonlinear evolution with their porosity, the microstructural origins of which are still questioning. Thus, the construction of the aforementioned links constitutes a subject of intense research. In particular, there is still a need for relevant multiscale and multiphysics models that could (1) account for the complexity of composite felt microstructures and related transport and sound absorbing properties and (2) be implemented in numerical simulation tools for computer-aided design of composite felts applications or for monitoring of the composite felts manufacturing process itself.

For that purpose, numerous theoretical studies have been conducted in the last decades. In most cases, composite felts are modeled as aligned fiber bundles (Berdichevsky and Cai \cite{berdichevsky1993preform},  Boutin \cite{boutin2000study},  Thiery and Boutin 
 \cite{thiery2020static}, Piegay \textit{et al.} \cite{piegay2021self},  Tarnow \cite{tarnow1996airflow, tarnow1996compressibility, tarnow1997calculation}, Umnova \textit{et al.} \cite{umnova2009effect}, Semeniuk and Goransson \cite{semeniuk2017microstructure},  Semeniuk \textit{et al.} \cite{Semeniuk2019}). These approaches often assume that the representative elementary volume of a composite felt can be reduced to the most basic geometric information, that is, porosity $\phi$ and fiber size, so that it is based on a bicomposite cylindrical pattern made of an internal cylindrical fiber and an external fluid shell that ensures fluid connectivity. The proposed analytical models are interesting because they encapsulate the essential parts of the physics and are easily configurable. However, they do not account for the complexity of the geometry and the combined effect that spatial randomness in the pore space has on flow problems.

To better understand the effects of visco-thermal micro-mechanisms on the values of transport coefficients of composite felts, many fiber scale numerical studies were conducted (Koponen \textit{et al.} \cite{koponen1998permeability}, Martys and Garboczi \cite{martys1992length}, Tomadakis and Robertson \cite{tomadakis2003pore, tomadakis2005survival}, Schladitz \textit{et al.} 
 \cite{schladitz2006design}, Umnova \textit{et al.} \cite{umnova2009effect}, Altendorf and Jeulin \cite{altendorf2011random}, Peyrega \textit{et al.} \cite{peyrega20113d}, Luu \textit{et al.} \cite{luu2017three, luu2017influence}, He \textit{et al.} \cite{he2018multiscale}, Soltani \textit{et al.} \cite{soltani2014effect}, Tucny \textit{et al.} \cite{tucny2022impact}). For example, Luu \textit{et al.} \cite{luu2017three, luu2017influence} performed numerical simulations using networks of straight cylindrical fibers to investigate the effect of porosity, fiber radius, and fiber orientation on the in-plane and through-plane transport properties of  fibrous media. Using random porous media from two-dimensional models, Martys and Garboczi \cite{martys1992length} demonstrated the important effect that spatial randomness in the pore space has on flow problems. This analysis showed that, in a random pore structure with a distribution of pore sizes, the viscous fluid flow will tend to go through the largest pore necks, decreasing the importance of the narrowest necks. They also highlighted that the sizes of the dynamically connected pore regions were not exactly the same for the electric and fluid flow cases (Martys and Garboczi \cite{martys1992length}). In particular, for a fibrous material made of wood fibers with an open porosity $\phi = 0.64$, Peyrega and Jeulin \cite{Peyrega2013} and Peyrega \textit{et al.} \cite{peyrega20113d} showed that the volume-weighted average radius $r_v$ was an appropriate size of the fiber radii to quantitatively predict its sound absorbing properties at normal incidence. This approach assumed a two-dimensional Boolean model of random cylinders composed of overlapping fibers, where the locations of the centers of the discs were determined according to a random Poisson point process. This analysis was extended to three-dimensional models for glass wool samples obtained with various processing parameters by He \textit{et al.} \cite{he2018multiscale}.
 
 Keeping in mind that at fixed porosity, fewer fibers are introduced into a given volume when the fiber radius is volume-weighted, these results highlight that the $r_v$ length scale provides an effective way to reconstruct pore space. This space encompasses the largest pores, forming a continuous path for the flow of viscous fluids in actual fibrous media.  These numerical results confirm the trends reported in several complementary experimental and semi-empirical studies (Delany and Bazley \cite{delany1970acoustical}, Bies and Hansen \cite{bies1980flow}, Miki \cite{miki1990acoustical}, Garai and Pompoli \cite{garai2005simple}, Manning and Panneton \cite{manning2013acoustical},  Kerdudou \textit{et al.} \cite{kerdudou:hal-01163424}, Xue \textit{et al.} \cite{xue2018prediction}, Pelegrinis \textit{et al.} \cite{PELEGRINIS20161}).  Briefly, they highlight (i) the central role of fiber distributions (in size and orientation) and (ii) the need for a proper characterization of the geometry and transport processes in polydisperse fiber structures. This is particularly true for nonwovens that exhibit a wide distribution of fiber diameters and lengths \cite{peyrega20113d, he2018multiscale}.

On the one hand, noticeable progress has been achieved in the purpose of characterizing the transport parameters of porous materials thanks to dedicated testing devices (Stinson
and Daigle \cite{stinson1988electronic}, Leclaire \textit{et al.} \cite{leclaire1996determination},  Ayrault \textit{et al.} \cite{ayrault1999ultrasonic}). These tests are interesting but still remain difficult to carry out, as they require permeable porous media to enable the propagation of ultrasonic waves through the thickness of the material. On the other hand, significant progress has also been achieved to characterize finely the microstructures of nonwoven fibrous media with imaging techniques such as scanning electron microscope images (Luu \textit{et al.} \cite{luu2017three}) and optical granulomorphometry (He \textit{et al.} \cite{he2018multiscale}) or X-ray microtomography coupled with advanced image analysis procedures (Lux \cite{lux2005comportement}, Peyrega \textit{et al.} \cite{peyrega20093d},  Depriester \textit{et al.} \cite{depriester2022individual}). For instance, He \textit{et al.} investigated the effect of fiber distributions (orientation, length, diameter) on several transport parameters of low density glass wools from optical granulomorphometry (length, diameter) and scanning electron microscope images (orientation) for ten products provided with two different classes of surface densities. Angular orientation and volume averaging of fiber diameters were used to reconstruct virtual geometries and quantitatively predict the viscous permeabilities $k_0$ of the corresponding samples. However, they did not fully capture the overall transport properties, in particular with respect to the high frequency parameters (viscous $\Lambda$ and thermal $\Lambda'$ characteristic lengths). 

In light of the above, the objective of this study is to propose a multiscale model for the overall transport and long-wavelength sound-absorbing properties of composite felts, taking into account the appropriate descriptors of the polydisperse microstructure that can be obtained using images.  For this purpose, two types of composite nonwovens with several compression ratios were manufactured and thermobonded. Their microstructures were characterized using scanning electron microscope images. We also characterized their transport and sound-absorbing properties. The combination of these data makes it possible to formulate relevant hypotheses for the architecture of fiber networks and their transport processes on the fiber scale. These features were then upscaled using homogenization with multiple scale asymptotic expansions for periodic structures (Sanchez-Palencia \cite{sanchez1980fluid}, Bensoussan \textit{et al.} \cite{Bensoussan1978}, Auriault \textit{et al.} \cite{auriault1977etude}). This method proposes a rigorous framework to deduce the effective coefficients of importance and the effective equations that govern the macro-fields of the equivalent continuum media of composite felts. It also provides well-posed boundary value problems to be solved on representative elementary volumes (REVs) to estimate their macroscale properties. These problems are first solved numerically using the finite element method. Then, a second semi-analytical multiscale model is proposed, approximating the numerical results obtained by curve fitting and yielding unified models which assume the effective coefficients of importance as a function of porosity, fiber orientations, and effective fiber radii. Predictions of the numerical and semi-analytical models are compared with experimental data and discussed.

 \section{Materials and experimental methods}
\label{sec:Materials and experimental methods}
\subsection{Felts}
\label{subsec:Polydispersity fibers and related}
\begin{figure}[ht]
\centering
 \includegraphics[width=9cm]{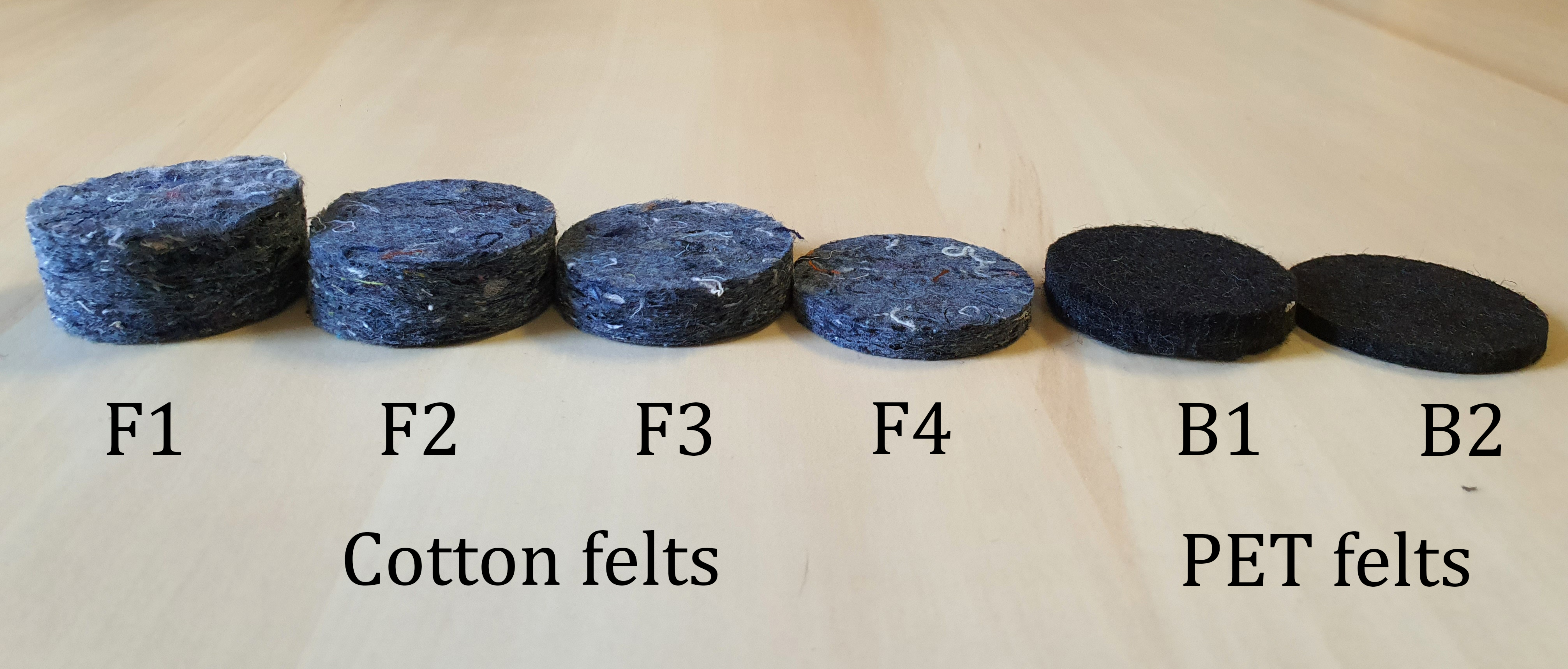}
 \caption{ Felt samples thermobonded at different thicknesses (sample diameter, 45$mm$).}
 \label{fig:felt}
\end{figure}

Two nonwoven materials are investigated (Fig. \ref{fig:felt}): namely ``cotton felt" and ``PET felt". Raw materials entering into the initial composition (Tab. \ref{tab:information fiber}) together with the corresponding manufacturing process are discussed in the following. Note that in the textile industry, the fineness $(t)$ of the fibers is specified by dtex, which enables a linear density estimate of the fiber size. To calculate the diameter of the fiber from the fineness $t$ (dtex) and the mass given by unit volume of the fiber material $\rho_f$, the following formula is used: $D_f =\sqrt{4t/\pi\rho_f}$.

\subsubsection{Cotton felt}
The fabrication of the cotton felt uses an airlay process, where the aerodynamic web forming is a dry procedure to form a web out of a wide variety of staple fibers. The fibers leave from a rotating drum into a turbulent air flow. Suctioning into a perforated moving conveyor belt or a perforated drum leads to the formation of a random three-dimensional web structure (Handbook of nonwovens, Chap. 4 \cite{BHAT2007143}; Gramsch \textit{et al.} \cite{gramsch2016aerodynamic}). \\
The input fiber material is a mixture of 75$\%$ shoddy fibers and 25$\%$ bicomponent fibers in mass.  The core of the bicomponent is made of PET, and its surface is made of coPET in a 1:1 ratio.  The bicomponent fibers are homogeneous with circular cross sections, whereas the shoddy fibers obtained after tearing of textile waste are not homogeneous. This shoddy is made from a mixture of 55$\%$ cotton and 45$\%$ PET.  In post-processing, the nonwoven material called felt is reinforced by thermobonding with a chosen compression ratio. Here, the bicomponent fibers have an adhesive effect.
\subsubsection{PET felt}
The input fiber material is a mixture of 60$\%$ PET fibers and 40$\%$  bicomponent fibers in mass.  The same bicomponent fibers are used as for cotton felts. The fibers are homogeneous with circular cross sections and regular lengths.  In web forming, the web is formed by a roller card. Fiber tufts and bundles are disentangled to form a parallel layer of fibers. The fibers in the card web have a lengthwise orientation. Then, this card web is laid in several layers on a take-off belt via a conveyor belt system with an oscillating carriage movement. This take-off belt moves 90 degrees to the cross-lapper. The fiber web is mechanically bonded by needling through the use of barbed needles. A portion of horizontal fibers are reoriented into the vertical plane in the form of fiber tufts. This nonwoven material is called needlefelt (Handbook of nonwovens, Chap. 8 \cite{AHMED2007368}; Nonwoven Fabric, Chap. 6 \cite{albrecht2006nonwoven}). Finally, thermobonding reinforcement is also applied along with the chosen compression ratio. 

\begin{figure}[ht]
\centering
 \includegraphics[width=6cm]{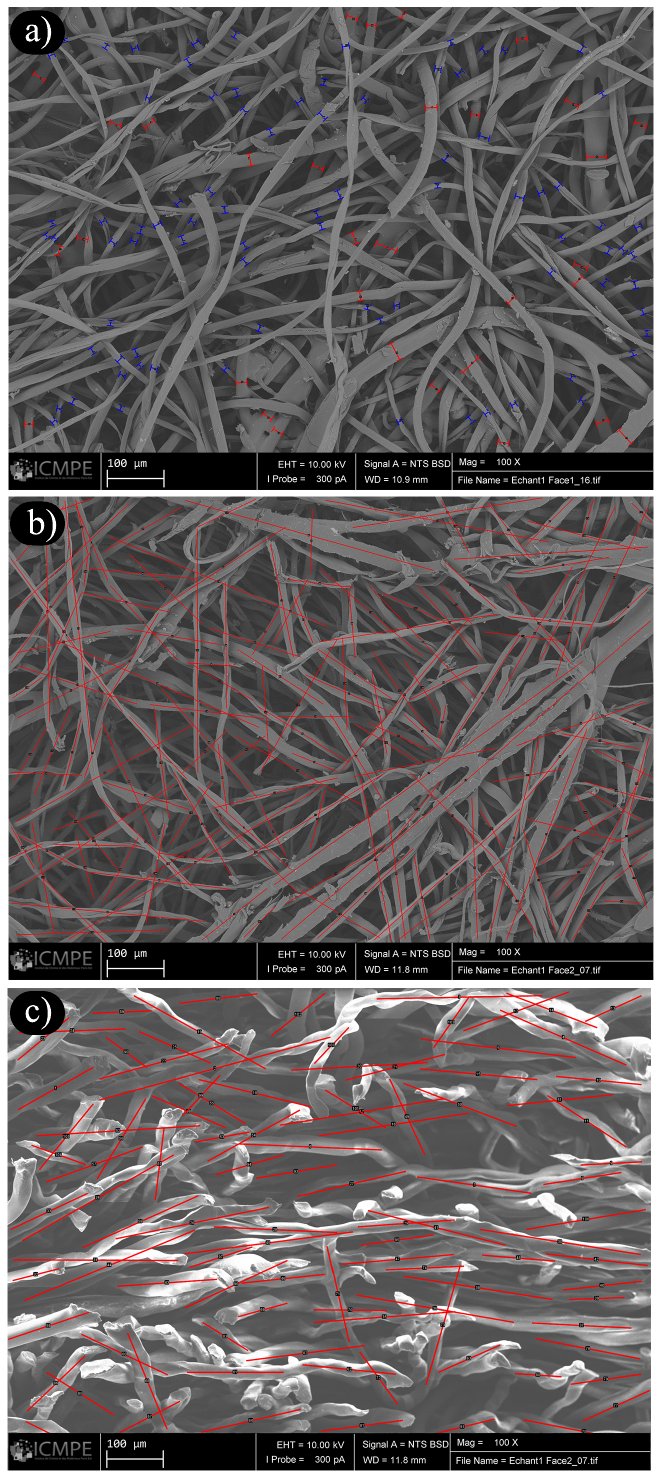}
\caption{Example of SEM images of cotton felt F2 and dimensional measurements of fibers (Fiji software). Measurement of: (a) fiber diameters (blue is cotton fibers, red is bicomponent fiber) in the $xy$-plane; (b) azimuthal or horizontal angle $(\varphi)$ in the $xy$-plane, and (c) zenithal or vertical angle measurements $(\theta)$ in the $xz$-plane.}
\label{fig:SEM}
\end{figure}

\begin{table*}[ht]
\centering
\begin{tabular}{ccccccc}
\hline\hline
Cotton felt & \multirow{2}{*}{\begin{tabular}[c]{@{}c@{}}Thickness \\ $(mm)$\end{tabular}} & Compression ratio & \multirow{2}{*}{\begin{tabular}[c]{@{}c@{}}Density\\  $(kg/m^3)$\end{tabular}} & \multicolumn{3}{c}{Mass composition}                                                                                                                                             \\ \cline{5-7} 
             &                                                                            &                  &                                                                                & Bicomponent                                                  & \multicolumn{2}{c}{Shoddy}                                                                                    \\ \hline
F1           & $20.3\pm0.3$                                                               & $1.0\pm0.00$          & $56.9\pm5.5$                                                                   & $25\%$                                                     & \multicolumn{2}{c}{$75\%$}                                                                                          \\
F2           & $16.1\pm0.6$                                                               & $1.3\pm0.05$     & $72.4\pm10.6$                                                                  & $25\%$                                                     & \multicolumn{2}{c}{$75\%$}                                                                                          \\
F3           & $11.2\pm0.2$                                                               & $1.8\pm0.04$     & $103.5\pm10.8$                                                                 & $25\%$                                                     & \multicolumn{2}{c}{$75\%$}                                                                                          \\
F4           & $5.9\pm0.2$                                                                & $3.4\pm0.10$      & $184.8\pm27.9$                                                                 & $25\%$                                                     & \multicolumn{2}{c}{$75\%$}                                                                                          \\ \hline
PET felt    & \multirow{2}{*}{\begin{tabular}[c]{@{}c@{}}Thickness\\  $(mm)$\end{tabular}} & Compression ratio & \multirow{2}{*}{\begin{tabular}[c]{@{}c@{}}Density\\  $(kg/m^3)$\end{tabular}} & \multicolumn{3}{c}{Mass composition}                                                                                                                                             \\ \cline{5-7} 
             &                                                                            &                  &                                                                                & \begin{tabular}[c]{@{}c@{}}Bicomponent \\ $4.4 dtex$\end{tabular} & \begin{tabular}[c]{@{}c@{}}PET\\  $6.7 dtex$\end{tabular} & \begin{tabular}[c]{@{}c@{}}PET\\ $17 dtex$\end{tabular} \\ \hline
B1           & $10.3\pm0.5$                                                               & $1.0\pm0.00$          & $141.0\pm8.4$                                                                  & $40\%$                                                     & $30\%$                                                    & $30\%$                                                  \\
B2           & $4.3\pm0.1$                                                                & $2.4\pm0.17$      & $344.9\pm15.5$                                                                 & $40\%$                                                     & $30\%$                                                    & $30\%$                                                  \\ \hline\hline
\end{tabular}
\caption{Information of the cotton felts and PET felts}
\label{tab:information fiber}
\end{table*}
\subsection{Characterization of the microstructure} 
The microstructure of the non-woven fibrous media was first characterized using Scanning Electron Microscope (SEM) images (Fig. \ref{fig:SEM}). The reader is referred to \ref{Appendix: Protocol} for a detailed description of the preparation and cutting of samples prior to acquisition of SEM images, and to \ref{Appendix: Calculation of the bias induced from the projection process} for a discussion on the bias introduced by the projection process. Based on these two-dimensional images, typical fiber diameters were measured manually (Figs. \ref{fig:SEM}a). To determine the in-plane [respectively out-of-plane] orientation distributions of the fibers, we superimposed straight segments on the fibers on the surface of the fibrous materials and extracted the in-plane orientation angle $\varphi$ (Fig. \ref{fig:SEM}b) [respectively out-of-plane orientation angle $\theta$ (Fig. \ref{fig:SEM}c)] for each segment of all identified fibers on orthogonal sections of the materials.

\subsection{Characterization of transport and acoustic properties} 
\label{sec:method for transport and acoustic}
The open porosity $\phi$ and true mass density  $\rho$ were determined using the pressure / mass method (Salissou and Panneton \cite{doi:10.1063/1.2749486}). This method makes it possible to precisely determine the uncertainty in porosity depending on the volume of samples tested. This is important since open porosity will be a fundamental property for the proposed multiscale model to work properly.  For each felt family (cotton and PET) and each fibrous material in a family (F1, F2, F3, F4, B1, B2), the density and porosity were measured. To ensure sufficient precision of measurements, for each fibrous material within a family, measurements were performed in batches of $12$ specimens of cylindrical samples with a diameter of $45$ mm (repeated three times per batch).

The airflow resistivity $\sigma$ was measured at a flow velocity of 0.5 mm/s following the static airflow method described in the ISO 9053-1:2018 standard. For each fibrous material, three cylindrical samples with a diameter of $45$ mm were cut and all leaks were carefully avoided by adding petroleum jelly to the circumference of the sample.

The torturosity $\alpha_\infty$ was measured using the high-frequency ultrasound transmission technique (Allard \textit{et al.} \cite{allard1994evaluation}). Three samples with a diameter of $100$ mm for each fibrous material were measured in air. 

The thermal characteristic length $\Lambda'$ also known as the generalized hydraulic radius is defined as two times the ratio between the fluid volume over wetted solid surface area of the porous material (Champoux and Allard \cite{champoux1991dynamic}). The viscous characteristic length $\Lambda$ was introduced by Johnson \textit{et al.} \cite{johnson1986new} as a dynamically connected radius of the porous structure by introducing a weighting of both the numerator and denominator of the generalized hydraulic radius by the squared velocity of a non-viscous fluid. The viscous $\Lambda$ and thermal $\Lambda^\prime$ characteristic lengths could not be validly measured with the two-gas ultrasound transmission technique (air and argon, Leclaire \textit{et al.} \cite{leclaire1996determination}).  It was also impossible for us to obtain valid results with the acoustic method of Panneton and Olny (\cite{panneton2006acoustical}, \cite{olny2008acoustical}). Indeed, due to acoustic measurements limited to 4000 Hz and to vibration effects, the stationarity criterion of these methods was not respected over the characteristic lengths. The same was true for thermal static permeability $k_0^\prime$.  Consequently, the Kozeny-Carman formula approach, as described in Henry \textit{et al.} \cite{Henry1995}, was used to estimate the two characteristic lengths $\Lambda$ and $\Lambda'$. This approach involves using the directly measured values for porosity $\phi$, resistivity $\sigma$, and tortuosity $\alpha_\infty$, as detailed in \ref{Appendix: Estimation characteristic lengths}. 

For the same reason, only an estimate of the static thermal permeability $k'_0$ could be obtained. It used the following relation between $\Lambda'$ and $k'_0$ \cite{olny2008acoustical}:
\begin{align}
    k'_0=M'\frac{\phi\Lambda^{'2}}{8}. \label{eq:k0p estimation}
\end{align}
The coefficient $M'$ is the dimensionless thermal shape factor. It differs from unity when the shape of the porous medium does not consist of circular cylindrical pores arranged in a parallel formation. From an educated guess based on the mean value of the results found for fibers in Tab. II of \cite{olny2008acoustical}, it was set to $M'=2.09$.   Therefore, an estimate of $k'_0$ was obtained from this equation using the measured porosity and the estimated thermal characteristic length. 

Finally, the sound absorption coefficient (hard-backed) of each felt was measured at normal incidence in an acoustic impedance tube of $44.44$ mm in diameter. The incident acoustic plane wave traveled along the $z$-axis and excited the front (or rear) face of the felt in the $xy$-plane (refer to Fig. \ref{fig:SEM}). The three-microphone method described in the ISO 10534-2:2023 standard was used.  The microphone spacing and tube diameter allowed valid measurements in the frequency range $45$ to $4300$ Hz. Three samples per felt were measured on both faces to capture variations from one specimen to another and to verify how symmetric the felts were in thickness.  The side of the specimen that is not facing the sound excitation is in contact with a hard reflective backing. To prevent air leakage between the tube wall and the specimens, a thin layer of Teflon was applied around the sample.

\section{Experimental results and discussion} \label{sec:Experimental results and discussion}
\subsection{Characterization of the microstructure}
The SEM images shown in Fig. \ref{fig:SEM} give typical features of the studied nonwoven fibrous media, fibers and fiber connections. From these images and the corresponding measurements, several important remarks can be made.

\begin{figure*}[ht]
\centering
\includegraphics[width=14cm]{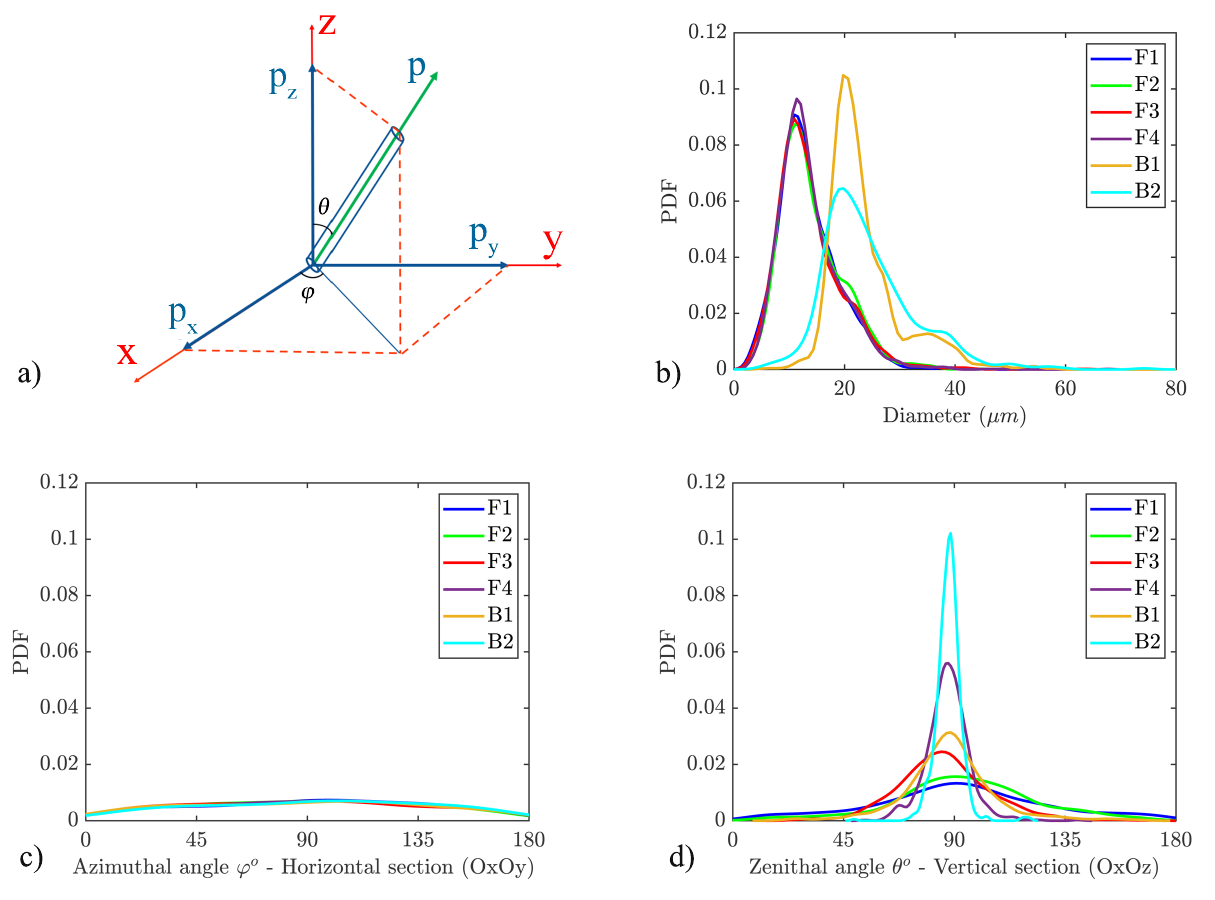}
\caption{(a) The orientation of a fiber in three-dimensional space in spherical coordinates. The estimated probability density functions of (b) the fiber diameter; (c) the azimuthal angle $\varphi$; (d) the zenithal angle $\theta$ as plotted using a non-parametric kenel method.}
\label{fig:diameter_angle} 
\end{figure*}
\begin{table*}[ht]
\centering
\begin{tabular}{ccccc}
\hline\hline
        & \multicolumn{2}{c}{Diameter} & \multicolumn{2}{c}{Zenith angle $\theta ^o$}\\ \hline
Samples & Number of measurements  & $D_m (\mu m)$ & Nb of measurements          &     $\theta ^o$ \\ \hline
F1      & $2386$                & $13.5\pm{5.6}$   & $823$                  & $91.9\pm36.2$ \\
F2      & $2086$                & $14.1\pm{5.8}$  & $850 $                 & $94.1\pm28.9$  \\
F3      & $2389$                & $13.8\pm{6.2}$  & $803$                  & $85.9\pm18.9$ \\
F4      & $2214$                & $13.7\pm{5.6}$  & $864$                  & $87.3\pm8.6$  \\
B1      & $2131$                & $23.6\pm{6.9}$  & $727 $                 & $87.9\pm18.6$ \\
B2      & $1780$                & $24.3\pm{8.4}$  & $644$                  & $87.7\pm5.3$ \\ \hline\hline
\end{tabular}
\caption{Statistics related to fiber diameters and angular orientation of fibers as experimentally determined from SEM images}
 \label{tab:diameter}
\end{table*}

\subsubsection{Fiber network}
Figure \ref{fig:SEM} shows that the nonwoven fibrous medium consisted of a more or less densely connected fibrous network through the heat bonding process. It shows a generally uniform fiber orientation distribution $\varphi$ in the $xy$-plane (Fig. \ref{fig:diameter_angle}c). The standard deviation on the out-of-plane angle $\theta$ decreases as the compression ratio increases (Fig. \ref{fig:diameter_angle}d, Tab. \ref{tab:diameter}). For all compression ratios (from F1 to F4 and B1 to B2), the average value of $\theta$ remains close to $90^o$.  These features reveal a transversely isotropic fiber orientation (see the corresponding second-order fiber orientation tensor in Advani and Tucker \cite{advani1987use}), which could be obtained using the numerical generation process parameterized with a preferred fiber alignment along the $Oz$ direction. Moreover, the observation regarding fiber connections tends to show that fibers can intersect; which could be considered in further simulations. 

\subsubsection{Fibers}
Figure \ref{fig:SEM} reveals that the fibers exhibited a rather small radius of curvature at the scale of a few hundred micrometers so that each fiber $i$ could generally be ascribed a mean tangent unit vector $\overrightarrow{p_i}$ to characterize its orientation (Fig. \ref{fig:diameter_angle}a). Furthermore, the fibers exhibited a more or less cylindrical shape with possible intersections due to the manufacturing process (Fig. \ref{fig:SEM}c). The fibers had a mean diameter $D_m=13.78$ $\mu$m for the cotton felts (F1-F4) and $D_m=23.95$ $\mu$m for the PET felts (B1-B2), see Tab. \ref{tab:diameter}. For each family of felts, these parameters were practically constant regardless of the compression ratio. In addition, the fiber diameter distributions were nearly the same for the cotton felt family (F1-F4), Fig. \ref{fig:diameter_angle}b. Finally, a small peak can be distinguished at $D_m$ = 20.1 $\mu$m that corresponds to the bicomponent fibers and a second peak at $D_m$ = 39.5 $\mu$m that corresponds to the second population of PET fibers (17 $dtex$).  The first population of PET fibers (6.7 $dtex$) does not appear clearly due to the fact that it is embedded in the central peak of PET felts. 

It should be mentioned that the thermocompression process on PET fibers had the effect of spreading the distributions of fiber diameter populations (Fig. \ref{fig:diameter_angle}b, B1, and B2).  This was not expected.  This may be because B1 was not heat-bonded, unlike B2, and the fibers were deformed in B2 after heat-bonded. Also, additional inaccuracy can be attributed to the manual measurement procedure. However, the two fiber diameter distributions were relatively similar at the end.

\subsection{Characterization of the transport properties}
\label{sec:results transport properties}
The transport properties, including the open porosity $\phi$, static airflow resistivity $\sigma$ (or alternatively static viscous permeability $k_0=\eta/\sigma$, where $\eta$ is the dynamic viscosity of the air), tortuosity $\alpha_{\infty}$, viscous $\Lambda$ and thermal $\Lambda'$ characteristic lengths, and static thermal permeability $k'_0$, are expected to be predicted using analytical expressions as a function of the morphological parameters. For example, Tarnow \cite{tarnow1996airflow} proposed an equation to determine the airflow resistivity for 2D cylinders of equal radii distributed in a square or random lattice.  Modifications of Tarnow's equations were suggested by Xue \textit{et al.} \cite{xue2018prediction} for situations in which a fibrous medium comprises more than one fiber component and when the radius of each fiber component varies within a certain range. Furthermore, Tamayol and Bahrami \cite{tamayol2010parallel} used a scale analysis technique (or semi-empirical approach) to determine the transverse permeability of various fibrous matrices, including square, staggered and hexagonal arrangements of aligned fibers, as well as simple two-directional mats and simple cubic structures. Umnova \textit{et al.} \cite{umnova2009effect} proposed an analytical method to predict the tortuosity, the characteristic lengths, and the static thermal permeability of a regular array of rigid parallel cylinders parallel or perpendicular to the flow direction. Pompoli and Bonfiglio \cite{pompoli2020definition} provided a modification of existing formulations of transport parameters based on numerical simulations for two-dimensional random structures considering fiber diameters with symmetric and asymmetric distribution. Luu \textit{et al.} \cite{luu2017influence} proposed a microstructural model for the transport parameters of three-dimensional networks of rectilinear fiber with constant diameter allowing for possible intersections. The equations were derived from rationalized numerical simulations in the form of master curves expressed as functions of porosity $\phi$, mean fiber radius $r_m$, and $\Omega_{zz}$ an effective parameter that parameterizes the angular orientation of the fibers.

For fibrous materials manufactured by thermo-compression with different thicknesses, Lei \textit{et al.} \cite{lei2018prediction} assumed that the transport parameters can be separated into two groups, depending ($\phi$, $\Lambda'$, $k'_0$) or not ($\sigma$, $\alpha_{\infty}$, $\Lambda$) on the orientation of the fibers. In their approach, porosity depends on the compression rate $n$ according to Castagnède \textit{et al.} \cite{castagnede2001parametric} formula; and $\Lambda'$ and $k'_0$ are determined as analytical functions of porosity $\phi$ (Umnova \textit{et al.} \cite{umnova2009effect}) as predicted by the Castagnède \textit{et al.} \cite{castagnede2001parametric} formula. Then, the model allowing prediction of $\sigma$ is an extension of the Tarnow \cite{tarnow1996airflow} model by considering averaging over an angular distribution function. The same principle is used to predict $\alpha_{\infty}$ and $\Lambda$, where this time the Umnova formula \cite{umnova2009effect} is used before performing the angular averaging. We note, however, that Lei \textit{et al.} \cite{lei2018prediction} model requires prior knowledge of the transport properties value before compression, which supposes available initial experimental measurements. 

\begin{figure*}[ht]
\centering
\includegraphics[width=14cm]{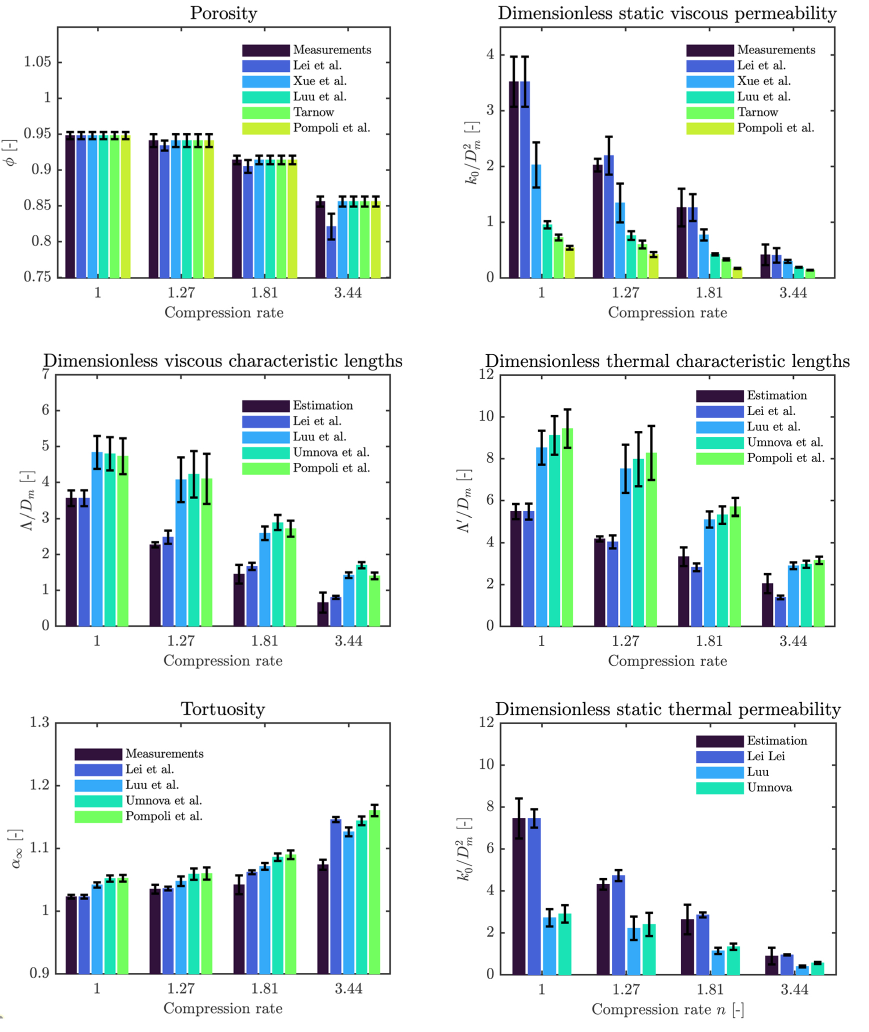}
\caption{Comparison between experimental estimates of the transport parameters on cotton felts F1 to F4 and the corresponding predictions with literature models (Lei \textit{et al.} \cite{lei2018prediction}, Xue \textit{et al.} \cite{xue2018prediction}, Luu \textit{et al.} \cite{luu2017three}, Umnova \textit{et al.} \cite{umnova2009effect}, Tarnow \cite{tarnow1996airflow}, Pompoli and Bonfiglio \cite{pompoli2020definition}). Note that the compression ratio of 1 refers to F1.}
\label{fig:compare_biblio} 
\end{figure*}

To compare the prediction of these models with our experimental data as a function of the compression ratio for cotton felts (F1-F4), we propose a standard dimensionless representation.  Here, the average fiber diameter $D_m$ is used to make all the dimensions of the transport properties dimensionless.  When fibrous materials are characterized by a wide distribution of fiber diameters (here the cotton-felt family, F1-F4), Fig. \ref{fig:compare_biblio} shows that the aforementioned models do not provide a relevant prediction for the transport parameters of the nonwoven fibrous materials studied. The model of Lei \textit{et al.} \cite{lei2018prediction} predicts the correct evolution of the transport parameters with the compression ratio when the experimental data are known at $n=1$. Hence, Fig. \ref{fig:compare_biblio} suggests that the transport behavior of the considered polydisperse fibrous media is ruled by representative volume elements different from those often assumed in previous models. In particular, we formulate the hypothesis that these models do not adequately account for the contribution of the polydispersity of fiber diameters and the particular physics induced by these geometries (Fig. \ref{fig:diameter_angle}b). 
\FloatBarrier
 \section{New microstructural model focusing on fiber characteristic sizes}
 \label{sec:A new microstructural model focusing on fibre characteristic sizes}
The experimental data collected in the previous section and the comparisons with literature models showed the difficulty of classical analytical microstructural models in  predicting the transport properties of nonwoven fibrous media with large fiber diameter polydispersity. Furthermore, these models do not always succeed in predicting transport properties as a function of the compression ratio.  Consequently, this section presents the development of two three-dimensional (3D) microstructural models in the porosity range $0.65\leq \phi\leq 0.99$.  These models take into account, in particular, the polydispersity of the fiber diameter and the fiber angular orientation. They are built on several assumptions related both to the fibrous microstructures and to fiber-scale thermoviscous dissipation mechanisms of locally heterogeneous fibrous materials in the long-wavelength regime. The approach assumes that the local characteristic sizes governing the transport phenomena within a polydisperse random fibrous microstructure depends on the time scale range of interest. This leads to the introduction of two specific diameters into the reconstruction procedure of two idealized 3D microstructures from which an upscaling technique is applied, namely the numerical homogenization method in the low- and high- frequency asymptotic regimes. Finally, additional equations are proposed to rationalize the results into compact analytical estimates for the dimensionless transport parameters of polydisperse fibrous structures.

\subsection{Idealized microstructures}
The typical Representative Elementary Volume (REV) of the nonwoven fibrous materials studied is seen as a 3D random fibrous network with $N$ straight cylindrical fibers.  A fiber $i$ in the REV is of diameter $D_i$ and defined by its center location $M_i$ and its orientation vector $\overrightarrow{p_i}$.  The fibrous medium studied exhibits a structure with transverse isotropy. Compressing the medium causes anisotropy. Following Schladitz \textit{et al.} \cite{schladitz2006design}, this anisotropy can be described by a density function of the directional distribution $p_{\beta}(\theta,\varphi)$ (Stoyan\textit{et al.} \cite{stoyan1980formulas}).  For the materials studied, with isotropy in the $xy$-plane, the function is:
\begin{align}
    p_\beta(\theta)=\frac{1}{4\pi}\frac{\beta sin(\theta)}{(1+(\beta^2-1)cos^2\theta)^\frac{3}{2}},\label{eq:pbeta}
\end{align}
where $\beta>0$ is the anisotropy parameter.

Furthermore, it is assumed that there is a good scale separation between the size $L$ of the reconstructed domain (REV size) and the smallest size between the macroscopic size of the nonwoven fibrous test samples (cylindrical samples with a diameter of $45$ mm) and the macroscopic size of the acoustic compression wave $\mathcal{L}=\lambda\/2\pi$ (of wavelength $\lambda$).  Note that the order of magnitude of size $L$ is given by the ratio $L/D_m$.  This magnitude is chosen so that the porosity of the REV is equal to the experimental value within $0.1\%$ of the relative difference. In addition, for the sake of simplicity, fibers are allowed to intersect during construction of a REV, which is consistent with the bounds visible on SEM images due to the thermo-compression process.

\begin{table}[ht]
\centering
\begin{tabular}{cccccc}
\hline \hline      
Samples  & $CV (\%)$       & $D_v (\mu m)$        &   $D_{iv} (\mu m)$   & $\beta $     & $\Omega_{zz} $ \\ \hline
F1       & $40.3$         & $19.5\pm{0.3}$        & $8.9\pm{0.4}$        & $1.4$        & $0.22$\\
F2       & $39.8$         & $18.7\pm{0.2}$        & $9.1\pm{0.3}$        & $1.7$        & $0.21$\\
F3       & $41.9$         & $19.1\pm{0.2}$        & $9.3\pm{0.2}$        & $3$          & $0.12$\\
F4       & $38.9$         & $18.5\pm{0.2}$        & $9.5\pm{0.2}$        & $6.5$        & $0.04$\\
B1       & $26.6$         & $26.7\pm{0.2}$        & $19.8\pm{0.3}$       & $3.5$        & $0.09$\\
B2       & $33.1$         & $31.2\pm{0.2}$        & $18.6\pm{0.3}$       & $12$         & $0.01$\\ \hline\hline
\end{tabular}
\caption{Estimated microstructural descriptors of the studied materials. $\Omega_{zz}$ is the angular orientation parameter \cite{luu2017influence}.}
 \label{tab:Pd_beta}
\end{table}

\subsection{Idealized transport phenomena}
In order to accurately upscale the transport and sound absorption phenomena of the fibrous media being studied, the diameters of the fibers are weighted according to their volume at low frequencies and inversely weighted according to their volume at high frequencies. These weighted diameters are given, respectively, by:
\begin{align}
    D_v=\frac{1}{\sum_{i=1}^{N_f}{V_i}}\sum_{i=1}^{N_f}{V_iD_i},\label{eq:Dv}
\end{align}
and
\begin{align}
    D_{iv}=\frac{1}{\sum_{i=1}^{N_f}{\frac{1}{V_i}}}\sum_{i=1}^{N_f}{\frac{1}{V_i}D_i},\label{eq:Div}
\end{align}
where $V_i$ is the volume of fiber $i$.  For the samples studied, these diameters are given in Tab. \ref{tab:Pd_beta}.

This apparently strong assumption is supported by the fact that the viscous boundary layer $\delta_v$ scales as $\sqrt{\eta/(\omega\rho_0)} $, in which $\eta$ is the dynamic viscosity of the fluid, $\rho_0$ is its density at rest and $\omega$ is the angular frequency of the sound wave. Indeed, due to the large viscous boundary layer $\delta_v$ at low frequencies and the local heterogeneities in the fiber network, the flow will tend to pass more through the largest pore necks. On the other hand, at high frequencies, inertial forces associated with fluid density dominate fluid motion, increasing the importance of the narrowest necks (see \ref{Appendix: Smallest channels in polydisperse fibrous structures}). The reader is also referred to Martys and Garboczi \cite{martys1992length} for a basic description of these transport phenomena supplemented by computer simulation studies. Consequently, when the considered nonwoven fibrous materials are subjected to a macroscopic long-wavelength plane compressional wave, the elementary transport parameters corresponding to the propagation of the sound wave through the materials are mostly influenced by the largest fibers at low frequencies and the smallest fibers at high frequencies. Note that more small-volume fibers can be introduced into a REV of fixed volume and porosity than large-volume fibers. Therefore, REV containing small-volume fibers will contain narrower constrictions than REV filled with large-volume fibers.

It should be noted that a volume-weighted average diameter was previously introduced by Peyrega \textit{et al.} \cite{peyrega20113d} and He \textit{et al.} \cite{he2018multiscale} to predict with success the permeability of heterogeneous fibrous materials. Here, we extend this idea to the inverse volume-weigthed average diameter.  Physically, $D_{iv}$ is thought to be the counterpart of $D_v$ to create the pore space that contains the smallest pores that form a continuous pathway through the fibrous polydisperse material in the high-frequency regime. It is easy to see that these arguments can be generalized to thermal effects.

\subsection{Theoretical upscaling}
Under an harmonic excitation, at angular frequency $\omega$, the local fluid velocity is governed by the linearized Navier-Stokes equations.  At low frequencies, the viscous drag forces dominate, and the Navier-Stokes equations simplify to the Stokes equations where the fluid is incompressible.  At high frequencies, inertial forces dominate, and there is a strong analogy between the inertial flow problem and the electrical conduction problem (Brown \cite{brown1980connection}, Johnson \textit{et al.} \cite{johnson1986new}). In this case, the Navier-Stokes equations can be replaced by the electric conduction problem. In this high-frequency analogy, the solid phase acts as an insulator and the fluid phase as a conductor (Johnson \textit{et al.} \cite{johnson1987theory}, Zhou and Sheng \cite{zhou1989first}). 

Therefore, using this analogy together with theoretical developments (Auriault \textit{et al.} \cite{auriault1977etude},  Lévy \cite{levy1979propagation}), from the homogenization method for periodic structures with multiscale asymptotic expansions (Bensoussan \textit{et al.} \cite{bensoussan2011asymptotic},  Sanchez-Palencia \cite{sanchez1980non}), several interesting results can be mentioned. Among them, it is possible to show that the macroscopic transport properties of interest $(k_0; \alpha_{\infty}, \Lambda)$ derive from generic boundary value problems (Stokes problem; electric conduction problem). Furthermore, an approximate but robust function $k(\omega)$ can be provided that predicts the dependence of visco-inertial effects using the low $(k_0)$ and high $(\alpha_{\infty}, \Lambda)$ frequency properties as input to the model. Finally, an analog frequency-dependent description $k'(\omega)$ of the thermal exchanges between the frame and the saturating fluid involving two macroscopic transport properties $(k'_0; \Lambda')$ can also be introduced (Lafarge \textit{et al.} \cite{lafarge1997dynamic}).

\subsection{Estimates of the transport properties}
\subsubsection{Numerical homogenization} \label{subsubsection: Numerical homogenization}
Taking advantage of the analogy mentioned above and theoretical developments, the transport properties of the random fibrous microstructures of the model were determined using a finite element method to solve Stokes, Laplace, and Poisson equations in the pore space. The transport properties of the nonwoven fibrous materials are then calculated by (i) generating for each studied nonwoven fibrous material two REVs, one for each asymptotic regime; (ii) solving the local partial differential equations which govern the phenomena at low and high frequencies, and (iii) computing the resulting transport parameters thanks to spatial averaging of the resulting fields. 

\begin{figure*}[ht]
\centering
\includegraphics[width=14cm]{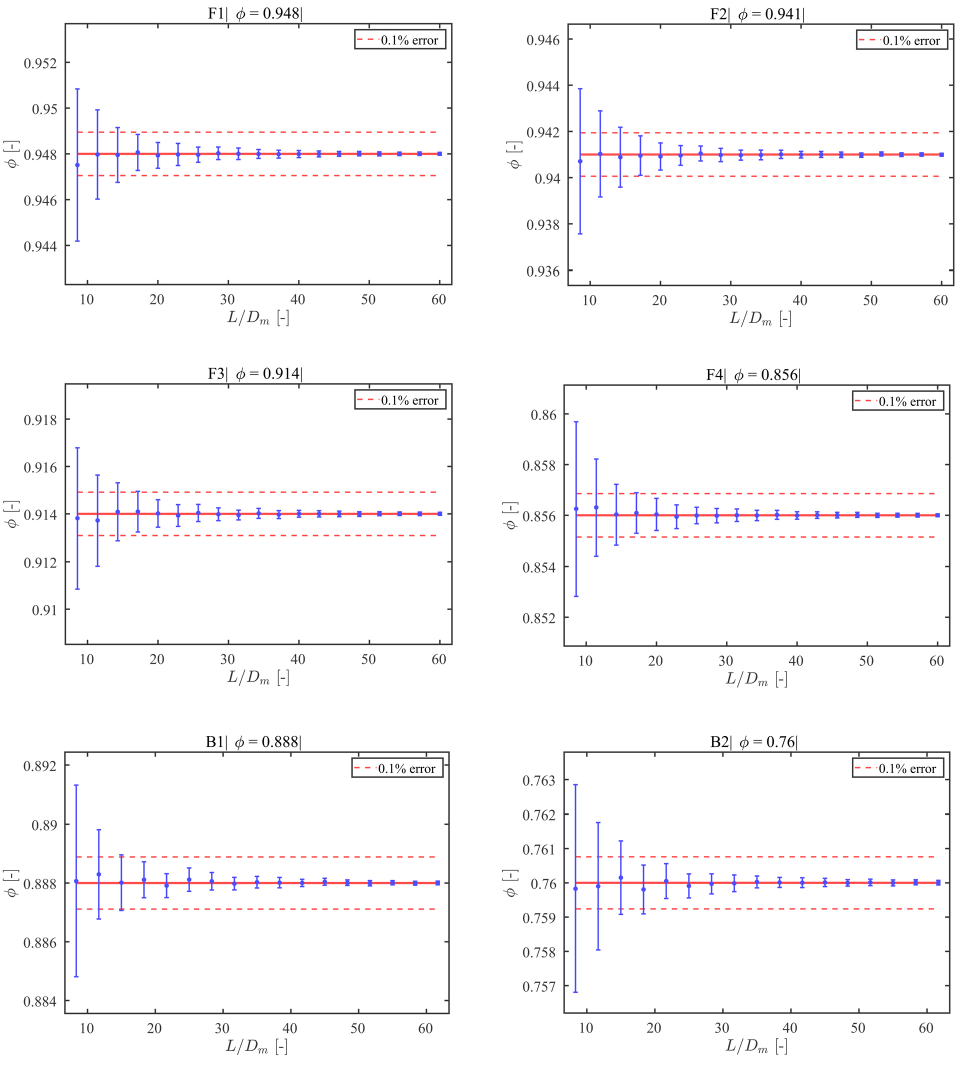}
\caption{\label{fig:domainsize} Evolution of porosity $\phi$ of the simulated three-dimensional random fibrous microstructures as a function of the size of the cubic box $L/D_m$, and comparison with the characterized value of porosity.}
\end{figure*}

For step (i), two series of numerical REVs, one with a mean volume-weighted diameter $D_v$ and one with a mean inverse volume-weighted fiber diameter $D_{iv}$, were generated to mimic the fibrous microstructures of the manufactured nonwoven seen by the sound wave in the low- and high-frequency regimes, respectively. Briefly, for each series, $N$ straight fibers $i$ of diameter $D_i$, with orientation vector $\overrightarrow{p_i}$, were generated within many REVs of volume $L^3$. Following Schladitz \textit{et al.} \cite{schladitz2006design}, Altendorf and Jeulin \cite{altendorf2011random}, Chapelle \textit{et al.} \cite{chapelle2015generation}, a stationary Poisson line process is defined with a one-parametric directional distribution $p_{\beta}(\theta,\varphi)$.  This parameter captures the degree to which the nonwoven is pressed. Practically, the values of $L$ were set such that the relative difference between the porosity of the geometric model and the measured porosity of the corresponding nonwoven fibrous material is less than $0.1\%$; for $100$ realizations of the geometrical model.  Figure \ref{fig:domainsize} presents a convergence study on $L$, for the materials studied, in terms of ratio $L/D_m$.  One can note that a ratio greater than 20 meets this porosity requirement. 

\begin{figure*}[ht]
\centering
\includegraphics[width=12cm]{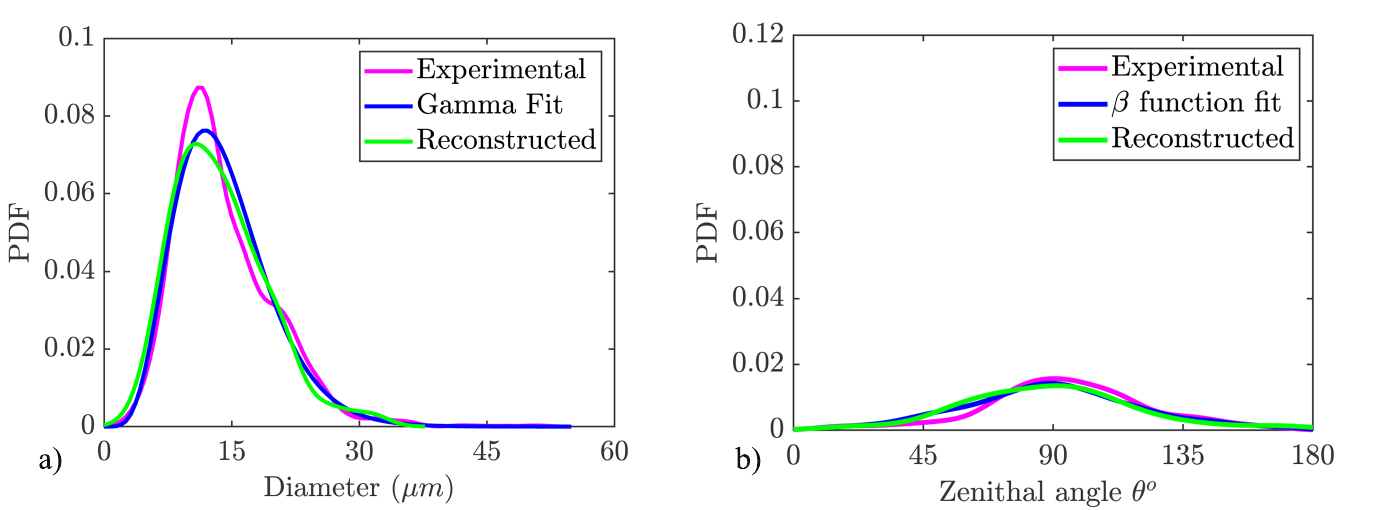}
\caption{Illustration of a comparison between the distributions of fiber diameters and  orientations as determined experimentally and from the corresponding models, also shown are the distributions after reconstruction.}
\label{fig:fit_beta_diameter}
\end{figure*}

\begin{figure*}[ht]
\centering
\includegraphics[width=12cm]{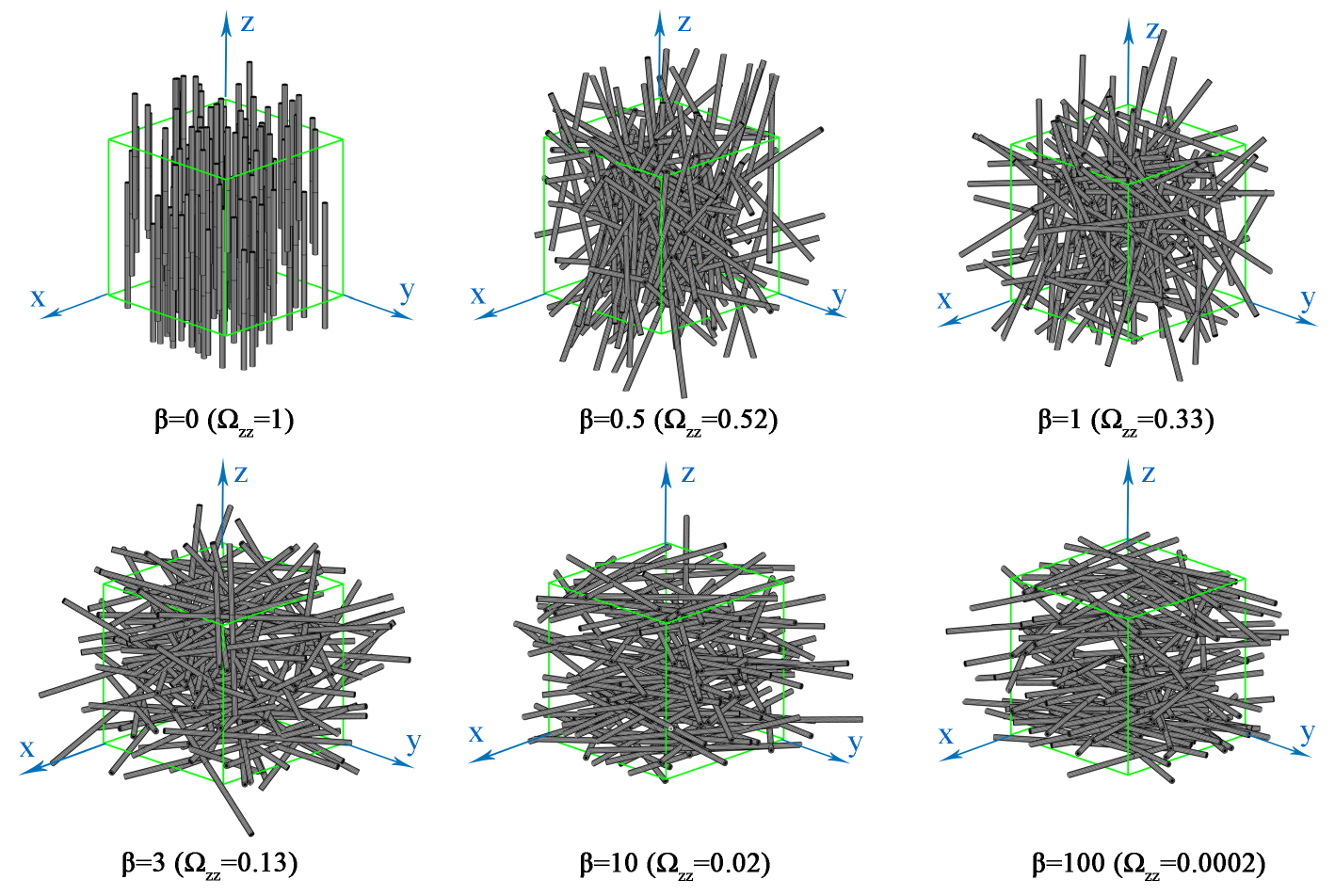}
\caption{Various configurations corresponding to the variation of fiber orientation states with $\beta$ ranging from 0 to 100, respectively.}
\label{fig:Pd_beta}
\end{figure*}

Thus, fiber networks were generated in REVs with various porosities $\phi$, ranging from $0.76$ to $0.948$, a fiber diameter distribution based on Gamma law, and a density function of directional distribution $p_{\beta}(\theta,\varphi)$, see Eq.\ref{eq:pbeta}.  The Gamma law and the parameter $\beta$ were determined by fitting the experimental results obtained from the SEM images.  An example for material F2 is shown in Fig. \ref{fig:fit_beta_diameter}. The figure shows the best-fit Gamma law and directional distribution.  The figure also shows that the generation procedure allowed fibrous networks to be obtained with fiber diameter and orientation distributions close to those measured experimentally. The best-fit $\beta$ values for each material samples are given in Tab.\ref{tab:Pd_beta}.  Figure \ref{fig:Pd_beta} shows six examples of idealized monodisperse fibrous networks with isotropic (or un-compressed) $(\beta=1)$, strechted ($\beta=0$) and compressed $(\beta>1)$ structures to show the influence of the parameter $\beta$.

For step (ii), periodic boundary conditions were ascribed to solve the boundary value problems on a REV. For a given fiber in contact with a couple of bounding surfaces, a point of the fiber was randomly determined along its length.  The fiber was cut at this point so that one segment of the fiber could be translated to maintain continuity at the boundaries.  A visual description of this process is given in \ref{Appendix:Geometrical reconstruction}.  

Figure \ref{fig:network} shows the periodic microstructural models reconstructed for material F2. Figure \ref{fig:network}a shows detailed information on the polydispersity of the fiber diameters and the directional distribution that accounts for the compression ratio. Three monodisperse models of the same medium are also presented: one with a mean fiber diameter $D_m$ (Fig. \ref{fig:network}b), one with a volume-weighted average diameter $D_v$ (Fig. \ref{fig:network}c)), and one with an inverse volume-weighted average diameter $D_{iv}$ (Fig. \ref{fig:network}d).

Assuming that all diameters follow a Gamma law, the polydispersity is easily quantified by the coefficient of variation $CV$.  This coefficient is defined as the ratio of the standard deviation on the fiber diameters to the mean value $D_m$.   For the materials studied, the values of $CV$ are given in Tab. \ref{tab:Pd_beta}.
Figure \ref{fig:network} underlines the inequality $D_v \geq D_m \geq D_{iv}$ and the interest in using the two different microstructural descriptors $D_v$ and $D_{iv}$ to predict the transport properties corresponding to, respectively, low-frequency $(k_0$, $k_0')$ and high-frequency ($\Lambda$, $\Lambda'$, $\alpha_{\infty}$) transport phenomena at known porosity.

\begin{figure*}[ht]
\centering
\includegraphics[width=14cm]{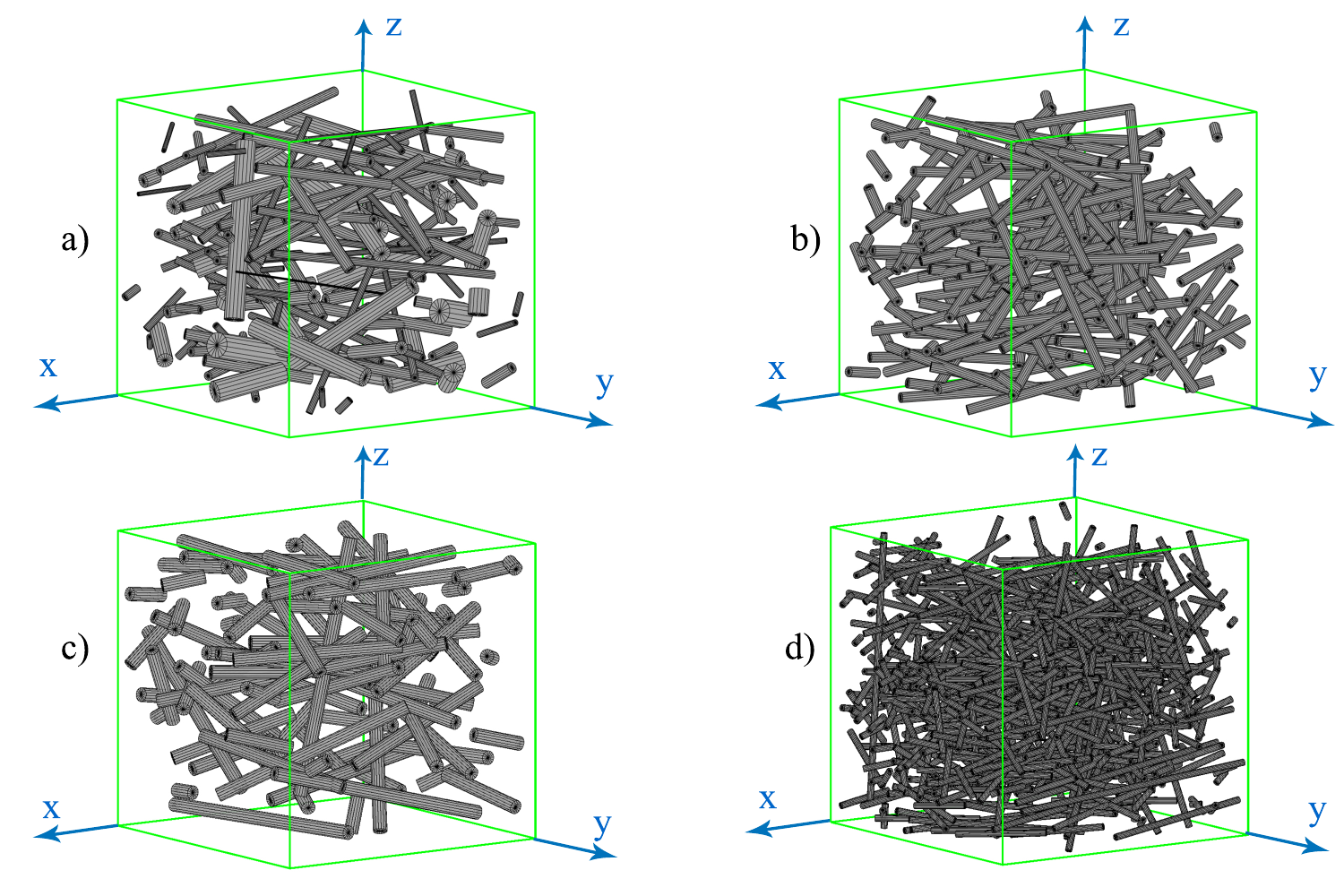}
\caption{\label{fig:network} Randomly overlapping fiber periodic structures of cotton felt F2; (a) polydisperse fibrous media; (b) monodisperse fibrous media with mean fiber diameter, $D_m$; (c) monodisperse fibrous media with volume-weighted mean diameter, $D_v$; (d) monodisperse fibrous media with inverse volume-weighted mean diameter, $D_{iv}$.}
\end{figure*}

\subsubsection{Semi-analytical model}
To build compact analytical expressions for the transport properties of nonwoven fibrous materials with relevant microstructural parameters, additional assumptions were stated on both the fibrous microstructures and the expected structures of the laws. Their relevance was checked using the microstructure generator and finite element simulations in the next section.  The assumptions and main expressions of the semi-analytical model are detailed in the following.
\begin{enumerate}
    \item The Gamma distribution offers a proper description of the distribution of fiber diameters.  One characteristic of this fiber diameter polydispersity is the coefficient of variation $CV$.  
    \item A stationary Poisson line process with a one-parametric directional distribution $p_{\beta}(\theta,\varphi)$ captures the angular orientation of a transversely isotropic fibrous medium and the degree to which the nonwoven was pressed. 
    \item The model should capture the geometry of the samples for a wide range of possible porosities $(0.65\leq\phi\leq0.99)$ and anisotropic parameters $(0 \leq \beta \leq 20)$. 
    \item A systematic mapping can be found by simulations in realizations of the geometric model.  On the one hand, this mapping allows us to define $r_v$ and $r_{iv}$ as functions of $r_m$ and $CV$, which are easily measurable microstructure descriptors. Here $r$ stands for radius.  On the other hand, there is a mapping between the anisotropy parameter $\beta$ and the orientation tensor governed by $\Omega_{zz}$ (Tab. \ref{tab:analytical model}).
\item The fibers could intersect so that $\Lambda'/r_{iv}$, the dimensionless ratio of two times the pore volume $V_p$ to pore surface area $S_p$ divided by inverse volume-weigthed average radius, can be written as given by Luu \textit{et al.} \cite{luu2017influence}:
\begin{align}
   \frac{\Lambda'}{r_{iv}}=\frac{\phi}{1-\phi+c},\label{eq:lct_analytic}
\end{align}
where $c$ is a constant accounting for the effects of fiber intersections on this high-frequency property.
\item  Archie's law \cite{archie1942electrical} that relates porosity to tortuosity holds.  This law is given by:
\begin{align}
\alpha_\infty = (1/\phi)^{\gamma}, \label{eq:tor_Archie}
\end{align}
where $\gamma$ is a constant that can vary between porous materials. This relation is defined for a series of materials from the same formation or manufacturing process. The detailed information on the pore structure is contained in the exponent $\gamma$. Theoretical studies have shown that $\gamma$ depends on the shape of the structuring element. When the microstructure is modeled as being built up of straight cylinders with mainly different orientations, a variable exponent could be used to handle the details of the pore space taken as a function of the angular orientation ($\beta$ or $\Omega_{zz}$) 
\begin{align}
   \alpha_{\infty z}=(\frac{1}{\phi})^{Q(\Omega_{zz})},\label{eq:tor_analytic}
\end{align}
where $Q(\Omega_{zz})$ is function of angular orientation  ($\beta$ or $\Omega_{zz}$).
\item The relation between the characteristic lengths derived by Johnson \textit{et al.} \cite{johnson1986new} holds.  This relation can be written as
\begin{align}
\frac{\Lambda^\prime}{\Lambda} =1-\frac{\ln(\alpha_\infty)}{\ln(\phi)}, \label{eq:lct_lcv_johnson}
\end{align}
This relation holds for the felts studied in which the porosity decreases by uniform growth of the insulating (solid) phase into the pore space. With Eq.\ref{eq:tor_analytic}, the previous relation becomes
\begin{align}
\frac{\Lambda^\prime}{\Lambda} = 1 + P(\Omega_{zz}). \label{eq:lct_lcv_analytic}
\end{align}
In principle, the function of angular orientation is the same as the one of Eq.\ref{eq:tor_analytic} but its fitted values could fluctuate to try to compensate for the oversimplifications of Eqs. \ref{eq:tor_Archie} and \ref{eq:lct_lcv_johnson}.  This is why $Q(\Omega_{zz})$ is replaced by a new function $P(\Omega_{zz})$.
\item Several classical models aim to represent the dependence of permeability on the geometric characteristics of the fiber network. The most classical model is the Kozeny-Carman equation (see Eq. (16) of \cite{pelegrinis2016application}) given by:
\begin{align}
\frac{k_0}{r^2_v} = \zeta\frac{\phi^3}{(1-\phi)^2}, \label{eq:k0_kozeny}
\end{align}
where $\zeta$ is the Kozeny ``constant'' which depends on the particle shape and size forming the solid skeleton. It can be shown that the through-plane normalized permeability $k_0/r_v^2$ also depends on the fiber orientations ($\beta$ or $\Omega_{zz}$). Indeed, the ratio $k_0/r_v^2$ increases significantly for larger fiber alignment in the direction of the macroscopic pressure gradient. It is assumed that a simple expression to estimate the normalized permeability $k_0/r_v^2$ as a function of $\phi^3/(1-\phi+m)^2$ and fiber orientation $(\Omega_{zz})$ can take the form
\begin{align}
\log_{10}\left(\frac{k_{0z}}{r_v^2}\right) = A\log_{10}\left(\frac{\phi^{3}}{(1-\phi+m)^{2}}\right) + S(\Omega_{zz}), \label{eq:k0_analytic}
\end{align}
where $A$ and $S(\Omega_{zz})$ are parameters to be calibrated by simulation for obtaining a general form.
\item Because diffusion of heat does not provide any preferred direction (spatially uniform heating), static thermal permeability $k'_0$, normalized by the square of the volume-weighted fiber radius $r_v^2$, can generally be written as a function independent of fiber orientation. In addition, the relation between $k'_0$ and $\Lambda'$ was introduced in Eq. \ref{eq:k0p estimation}. Then, combining Eqs.\ref{eq:k0p estimation} and \ref{eq:lct_analytic}, the normalized thermal permeability as a function of the open porosity can be expressed as
\begin{align}
\frac{k'_{0}}{r_v^2} = m_{1}\frac{\phi^{3}}{(1-\phi+m_{2})^{2}}, \label{eq:k0p_analytic}
\end{align}
where $m_1$ and $m_2$ are calibration constants. It should be noted that this relation is normalized by the volume-weighted fiber radius $r_v$, as $k'_0$ is a low-frequency parameter. The value of $m_1$ accounts for the shape of the porous network, while $m_2$ may be different from $c$ as the effects of the fiber intersections may be different at low frequencies.
\end{enumerate}

Equations \ref{eq:lct_analytic}, \ref{eq:tor_analytic}, \ref{eq:lct_lcv_analytic}, \ref{eq:k0_analytic} and \ref{eq:k0p_analytic} form the semi-analytical model (or micromacro relationships) for transversely isotropic polydisperse nonwoven fibrous media. They depend only on the open porosity $\phi$, the angular orientation ($\beta$ or $\Omega_{zz}$), and the coefficient of variation $CV$.  The main equations of the semi-analytical model are summarized in Tab. \ref{tab:analytical model}, where the constants and polynomials were determined with the numerical results presented in the following section.  

\begin{table*}[ht]
\centering
\begin{tabular}{llll}
\hline  \hline     
Parameter                           & Equation                                                  & R-squared \\ \hline
(a) Thermal length       & $\frac{\Lambda'}{r_{iv}}=\frac{\phi}{1-\phi+0.00073}$  & 0.999     \\
(b) Viscous length       & $\frac{\Lambda'}{\Lambda}=1+P(\Omega_{zz})$                 &  0.996     \\
 (c) Tortuosity                          & $\alpha_{\infty}=(\frac{1}{\phi})^{Q(\Omega_{zz})}$          & 0.977      \\
(d)  Viscous permeability         & $\log_{10}{(\frac{k_0}{r_v^2})}=0.7501\log_{10}{(\frac{\phi^3}{(1-\phi+0.0038)^2})}+S(\Omega_{zz})$                                                                      &0.996   \\
  (e) Thermal permeability         & $\frac{k'_0}{r_v^2}=0.08\frac{\phi^3}{(1-\phi+0.0173)^2}$      & 0.988   \\ 
(f) Coefficient $\Omega_{zz}$         & $\Omega_{zz}=0.8564e^{-0.8927\beta}+0.02163$      & 0.981   \\ 
\multirow{2}{*}{(g) Weighted radii}         & $\frac{r_v}{r_m}=0.0002CV^2 -0.00014 CV+1.003$                  &  0.999\\
   & $\frac{r_{iv}}{r_m}=0.909e^{-\left(\frac{CV-4.742}{31.97}\right)^2} + 0.417e^{-\left(\frac{CV-42.37}{25.92}\right)^2}$& 0.999  \\
\multirow{3}{*} {(h) Polynomials}                 & $P(\Omega_{zz})=-0.158\Omega_{zz}^2-0.666\Omega_{zz}+0.925$   &  \\ 
& $Q(\Omega_{zz})=-0.0914\Omega_{zz}^2-0.341\Omega_{zz}+0.495$   &  \\ 
 & $S(\Omega_{zz})=0.1313\Omega_{zz}^2+0.1755\Omega_{zz}-1.13$      & \\\hline\hline
\end{tabular}
\caption{Semi-analytical model equations to predict the transport properties of a fibrous material.}
 \label{tab:analytical model}
\end{table*}
\FloatBarrier
\section{Model prediction and discussion}
\label{sec:Model prediction and discussion}
\subsection{Numerical results}
    \begin{figure*}[ht]
    \centering
    \includegraphics[width=14cm]{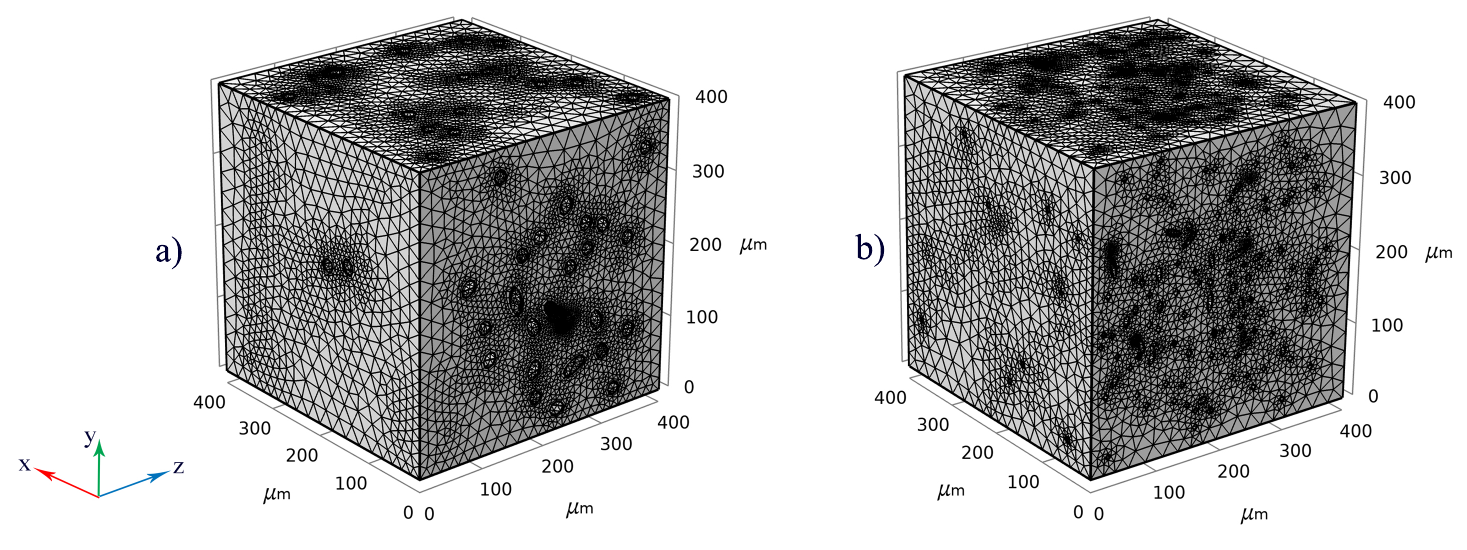}
     \caption{ Typical meshes of the fluid phase in a periodic REV of fibrous medium F2.  The meshes are used to perform finite element simulations on: a) structure with inverse volume-weighted diameter with 947,011 tetrahedral elements, and b) structure with volume-weighted diameter with 1,042,941 tetrahedral elements.}  
     \label{fig:mesh}
\end{figure*}
\begin{figure*}[ht]
\centering
\includegraphics[width=14cm]{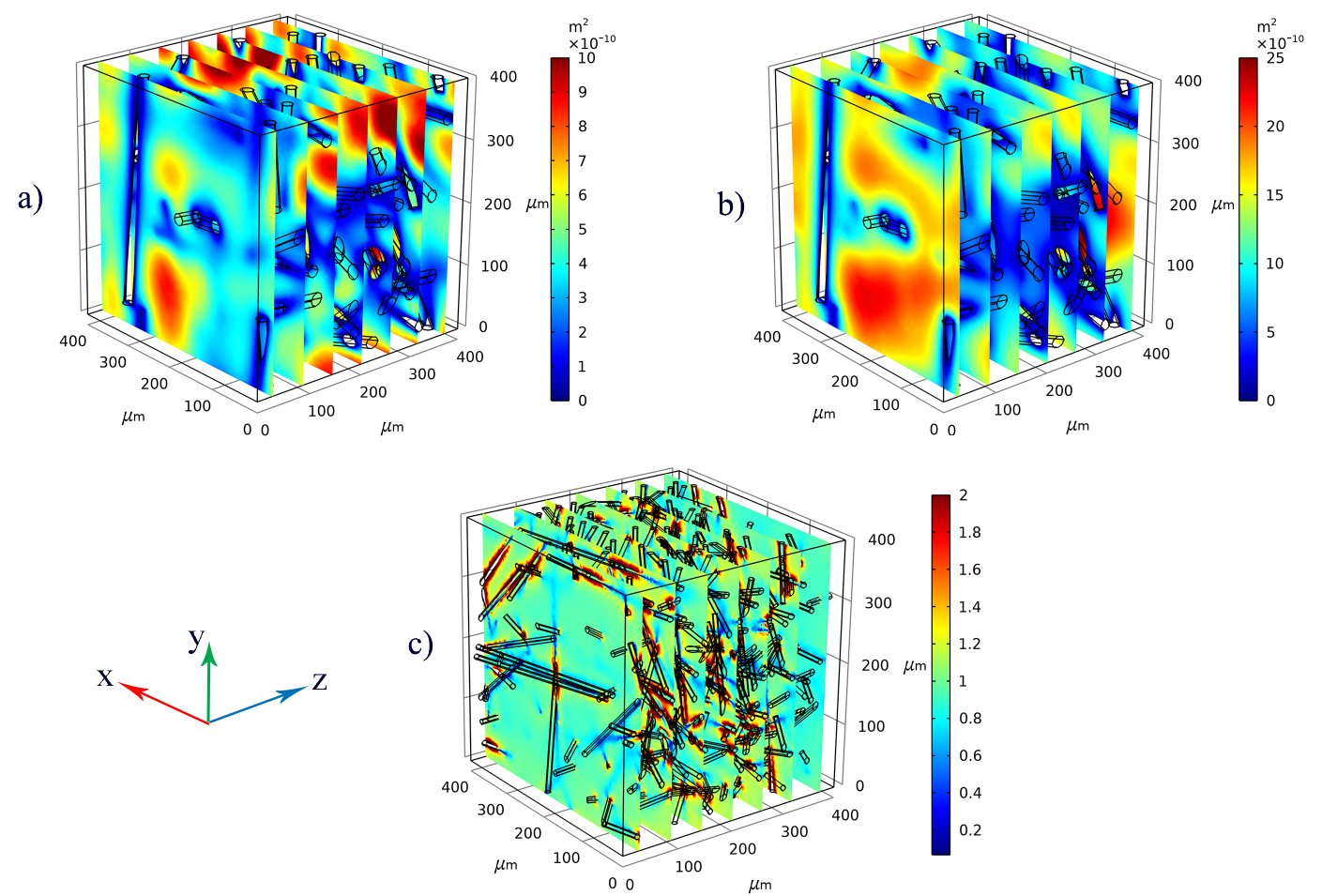}
\caption{Asymptotic fields of velocity and temperature computed on the discretized REVs of Fig.\ref{fig:mesh} for material F2: (a) scaled velocity field expressed as local permeability ($k_{0})$ $[m^2]$ corresponding to Stokes flow in the $z$ direction with the REV reconstructed by volume-weighted diameter; (b) scaled heat diffusion field expressed as local static thermal permeability $(k'_0)$ $[m^2] $ with the REV reconstructed by volume-weighted diameter, and (c) scaled velocity field expressed as tortuosity $\alpha_\infty$ $[-]$ corresponding to potential flow in the $z$ direction with the REV reconstructed by volume inverse weighted diameter.}
\label{fig:Solution} 
\end{figure*}
By taking advantage of two specific weighted fiber diameters, we have proposed that the studied polydisperse fibrous microstructures subjected to several compression rates and thermo-mechanical bounding could be modeled by two different REVs (i.e., volume weighted $r_v$ and inverse volume weighted $r_{iv}$ fiber radii) corresponding to the transport phenomena than can be simulated in the low and high frequency regimes. Therefore, it was possible to extract from three elementary boundary value problems (Stokes, Laplace, Poisson), and from the computation of the corresponding solution fields (Figs. \ref{fig:mesh} and \ref{fig:Solution}), the expressions of the through-plane static viscous $k_0$ and thermal $k'_0$ permeabilities of the nonwoven fibrous medium, as well as their through-plane viscous characteristic length $\Lambda$ and tortuosity $\alpha_{\infty}$. 
For its part, the thermal characteristic length $\Lambda'$ was calculated directly by twice the ratio of pore volume to surface area in each REV mesh. Figure \ref{fig:Para_anal_simu_fit}  shows the evolution of $k_0/r_v^2$, $k'_0/r_v^2$, $\Lambda/r_{iv}$, $\Lambda'/r_{iv}$, and $\alpha_{\infty}$ with the porosity $\phi$, for nonwovens with transverse isotropy and with a preferred orientation (Fig. \ref{fig:Pd_beta}). These predictions were obtained with a domain size $L/D$ allowing for convergence on porosity by taking five realizations for each porosity. From this figure, several remarks can be drawn:
\begin{itemize}
	 \item The through-plane viscous permeability $k_0/r_v^2$ increases non-linearly with the porosity and diverges as the porosity $\phi$ is approaching unity $(\sim \phi^3 / (1 - \phi)^2)$. At high porosities, the effect of preferred fiber orientation (induced by compression or manufacturing process) is strong and cannot be ignored. Lower viscous permeabilities are observed for in-plane fiber orientations than for out-of-plane fiber orientations, in agreement with previous results (Tarnow \cite{tarnow1996airflow}) . In contrast, the static thermal permeability $k'_0/r_v$ is independent of fiber orientation at a constant porosity. It is noteworthy that, as shown by the results of Fig. \ref{fig:Para_anal_simu_fit}a, the formal inequality $k'_0\geq k_0$ is also clearly apparent (Avellaneda and Torquato \cite{avellaneda1991rigorous}).
	 \item  Similarly, the viscous $\Lambda$ and thermal $\Lambda'$ characteristic lengths also increase non-linearly with the porosity $(\sim\phi/(1 - \phi))$ [Fig. \ref{fig:Para_anal_simu_fit}c-d] but to a lesser content than for the viscous $k_0$ and thermal $k'_0$ permeabilities  $(\sim\phi^3 / (1-\phi)^2)$ [Fig. \ref{fig:Para_anal_simu_fit}a-b]. We also checked the following inequality, $1 \leq\Lambda'/\Lambda\leq2$, available for fibrous media in the dilute limit $(\phi\rightarrow 1)$ \cite{champoux1991dynamic} ; with $\Lambda'/\Lambda= 1$ in the limit of in-plane orientation distributions of fibers and $\Lambda'/\Lambda= 2$ for fully aligned fibers (Fig. \ref{fig:Para_anal_simu_fit}e). This observation implies that the ratio $\Lambda'/\Lambda$ increases with the compression rate. The results of $\Lambda'/\Lambda$ were relatively independent of porosity [Eq. \ref{eq:lct_lcv_analytic}]. Increasing the fiber alignment significantly increases the viscous characteristic length [Fig. \ref{fig:Para_anal_simu_fit}c], the effect is larger for high porosities $(\Lambda' \sim \phi/(1 - \phi)$ and $\Lambda'/\Lambda = 1 + P(\Omega_{zz}))$ which occurs physically because $\Lambda$ is weighted by the scalar product of the local electric field solution $\textbf{E}\cdot\textbf{E}$ (in plane orientation of fibers creates smaller channels for the preferential fluid flow).
    \item The tortuosity $\alpha_{\infty}$ decreases with increasing porosity (Archie's law; $\alpha_{\infty}\rightarrow 1$  when $\phi\rightarrow 1$). However, apart from this limit $\phi\rightarrow 1$), the tortuosity $\alpha_{\infty}$ was shown to increase at constant porosity when the fibers are perpendicular to the potential flow direction. This situation corresponds to a more tortuous path (Fig. \ref{fig:Para_anal_simu_fit}f) for which a larger dispersion of the microscopic velocities is obtained (Eq. \ref{eq:tor}).
 \end{itemize}
\subsection{Comparison between finite element simulations and the semi-analytical model}
\begin{figure*}[ht]
\centering
\includegraphics[width=14cm]{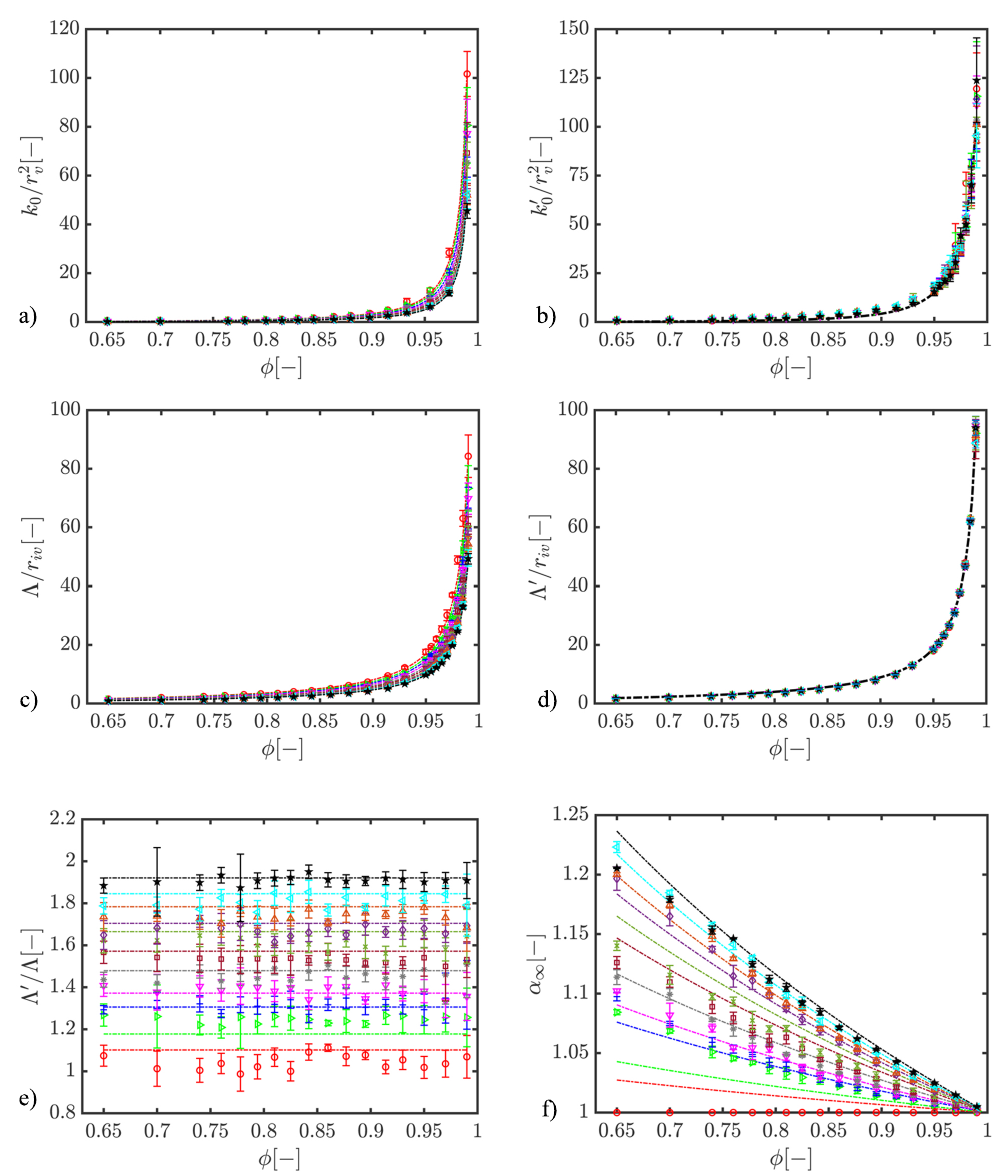}
\caption{\label{fig:Para_anal_simu_fit} Normalized transport parameters as a function of porosity $\phi$. The symbols indicate the statistically averaged orientation of fibers as determined by value of $\beta$ or $\Omega_{zz}$: $\Omega_{zz}=0$ (\textcolor{black}{$\star$}), $\Omega_{zz} = 0.11$ (\textcolor{Aqua}{$\triangleleft$}), $\Omega_{zz} = 0.19$ (\textcolor{orange}{$\triangle$}), $\Omega_{zz} = 0.30$ (\textcolor{violet}{$\diamond$}), $\Omega_{zz} = 0.39$ (\textcolor{YellowGreen}{$\times$}), $\Omega_{zz} = 0.49$ (\textcolor{FireBrick}{$\square$}), $\Omega_{zz} = 0.61$ (\textcolor{Grey}{$*$}), $\Omega_{zz} = 0.71$ (\textcolor{magenta}{$\triangledown$}), $\Omega_{zz} = 0.81$ (\textcolor{blue}{$+$}), $\Omega_{zz} = 0.91$ (\textcolor{green}{$\triangleright$}), $\Omega_{zz} = 1$ (\textcolor{red}{$\circ$}). The dashed lines are estimates obtained by the semi-analytical model derived from the numerical simulations (Tab. \ref{tab:analytical model}).}
\end{figure*}

Fig. \ref{fig:Para_anal_simu_fit} shows the five correlation functions $k_0/r_v^2$ $(\phi, \Omega_{zz})$, $k'_0/r_v^2$$(\phi)$,  $\Lambda/r_{iv}$ $(\phi, \Omega_{zz})$, $\Lambda'/r_{iv}$ $(\phi)$, $\alpha_{\infty}$ $(\phi, \Omega_{zz})$ with porosity, over a wide range of porosities $(0.65 \leq \phi \leq 0.99)$. The dashed lines in Fig. \ref{fig:Para_anal_simu_fit} are drawn from the results of curve fitting to the simulation data; the analytical expressions are those of Eqs. (\ref{eq:lct_analytic}, \ref{eq:tor_analytic}, \ref{eq:lct_lcv_analytic}, \ref{eq:k0_analytic}, \ref{eq:k0p_analytic})  and the fitted coefficients reported throughout Tab. \ref{tab:analytical model}. Good agreement with the simulation data is seen in all five cases, confirming the above initial law derivations. The square of the correlation coefficient gives R-squared $\geq 0.977$. The fact that R-squared is less than one indicates that at some porosities, there is a small proportion of Sum of Squared Errors (SSE) that is not accounted for by the regression. The coefficient of determination (R-squared) of the fit was 0.999 for the thermal characteristic length, 0.977 for the tortuosity, 0.996 for the ratio of the thermal over viscous characteristic lengths, 0.996 for the viscous permeability and 0.988 for the thermal permeability. The proportionate amount of variation in the response variable (dimensionless transport parameter) that is explained by the independent variables (porosity $\phi$ and orientation of fibers $\Omega_{zz}$) was therefore always very close to one. 

The residual analysis enables a local quantitative appreciation of the adequacy of the fitted model (Fig. \ref{fig:fitting tor lcvlct}). The residuals from a fitted model are defined as the differences between the response data (simulations) and the fitting to the response data (model) at each predictor value. The largest differences are obtained for the tortuosity $\alpha_{\infty}$, as $\phi \rightarrow 0.65$ and $\Omega_{zz} \rightarrow 1$. In this situation, the tortuosity values should correspond to the upper bound [Eqs. \ref{eq:tor_Archie} and \ref{eq:tor_analytic}] of a solid fibrous network with lower porosities. But if simulations are performed in opposite fiber orientations, from $\Omega_{zz}=0$ for in-plane fibers to $\Omega_{zz}=1$ for unidirectionally aligned fibers, a large variation of tortuosity values should be observed which is somehow contradictory with the initial choice of an Archie’s law [Eq. \ref{eq:tor_Archie}]. The presence of these contradictory behaviors ($\alpha_{\infty}$ increases with decreasing $\phi$, $\alpha_{\infty}$ decreases with increasing $\Omega_{zz}$) can be used to explain the higher sensitivity of the model to geometrical parameters and the larger proportion of numerical results not entirely present in the model. Similar arguments can be given to quantify the differences between the finite element simulations and the analytical model for the static thermal permeability $k'_0$: as $\phi \rightarrow 1$, the value of $k'_0$  diverges as $(\sim \phi^3 / (1 - \phi)^2)$ [Eq. \ref{eq:lct_analytic}] which statistically increases the proportion of SSE that is not completely explained by the regression. 

Our results suggest that a better fit would require an increase in the domain size $L/D_m$, which is important to ensure a lower relative difference between the porosities of the generated microstructures and the porosity that serves as the target, as $\phi$ approaches one. i.e., if $\phi$ target = 0.99 with err = 0.01$\%$, $ L/D_m = 55$; if $\phi$ target = 0.99 with err = 0.001$\%$, $L/D_m = 140$ (see Figs. \ref{fig:algo} and \ref{fig:domainsize}). 

Finally, in this section, we presented a comparison between the analytical result and the numerical finite element solution. We saw, through a detailed analysis of the residues, that the comparison between finite element simulations and the analytical model (Figs. \ref{fig:Para_anal_simu_fit} and \ref{fig:fitting tor lcvlct}, Tab. \ref{tab:analytical model}) revealed that the analytical expressions [Eqs. \ref{eq:lct_analytic}, \ref{eq:tor_analytic}, \ref{eq:lct_lcv_analytic} \ref{eq:k0_analytic}, \ref{eq:k0p_analytic}] fit well with the trends gained from the finite element simulation when the same microstructure parameters are used as input. Hence, analytical estimates can be considered to be accurate enough predictors of the transport properties of nonwoven fibrous materials.
\begin{figure*}[ht]
\centering
\includegraphics[width=14cm]{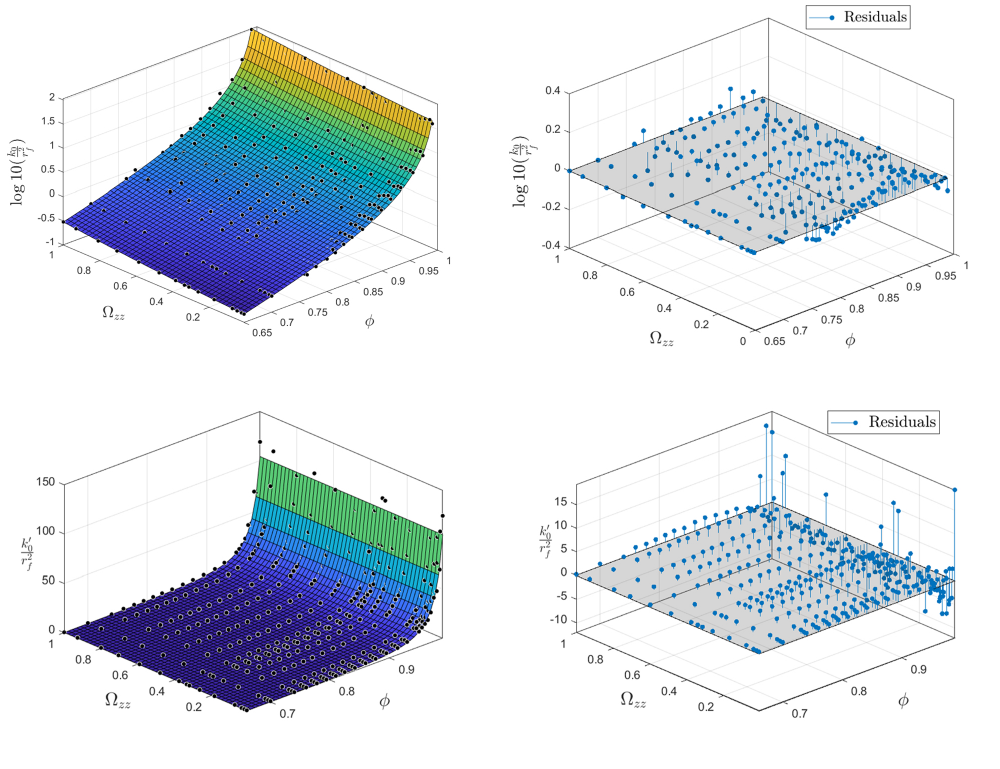}
\label{fig:fitting k0 k0p}
\end{figure*}
\begin{figure*}[ht]
\centering
\includegraphics[width=14cm]{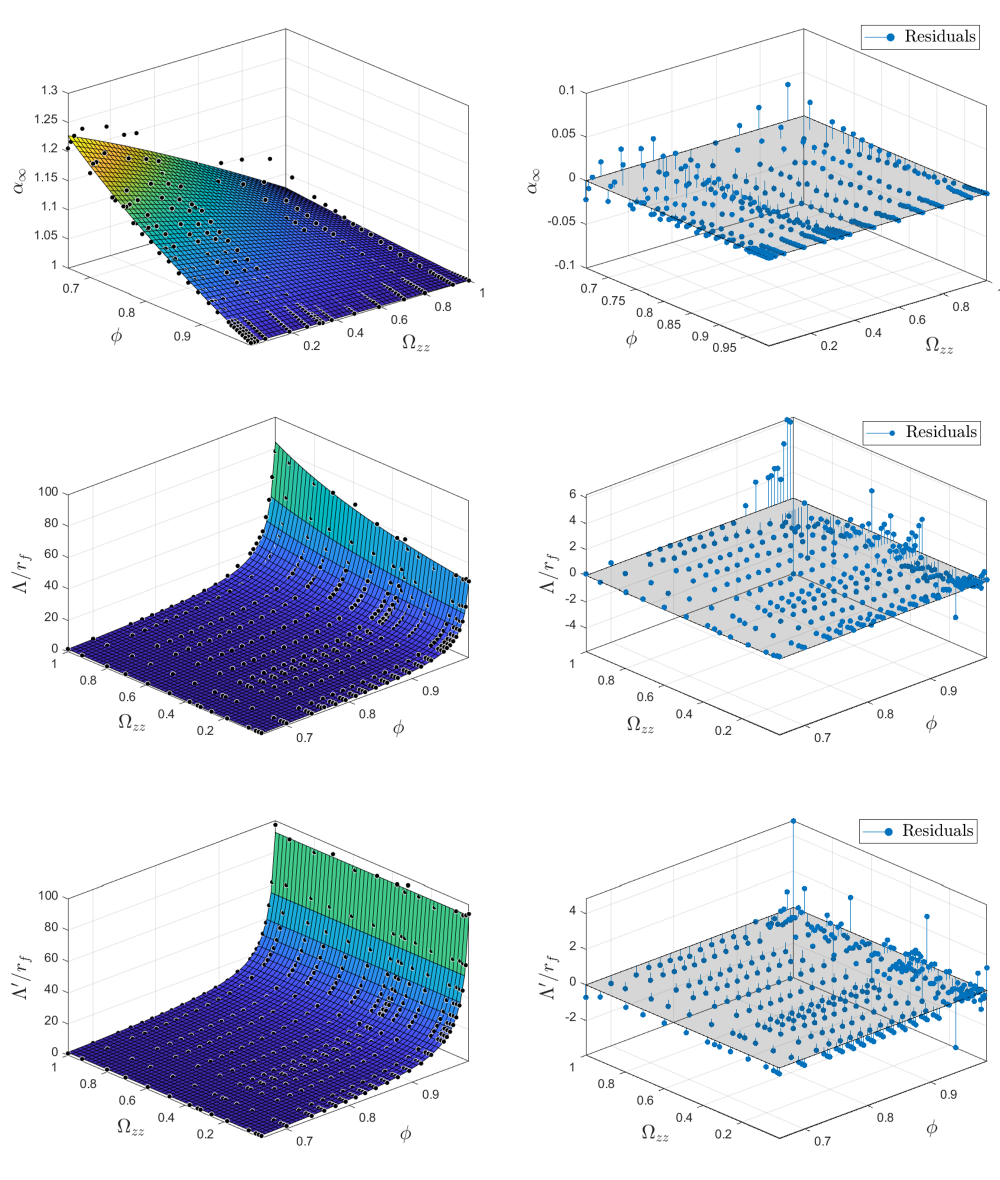}
\caption{\label{fig:fitting tor lcvlct} Map fittings and residual plots of dimensionless transport parameters (semi-analytical model derived from the numerical simulations).}
\end{figure*}
\subsection{Comparisons with experimental results}
\begin{figure*}[ht]
\centering
\includegraphics[width=14cm]{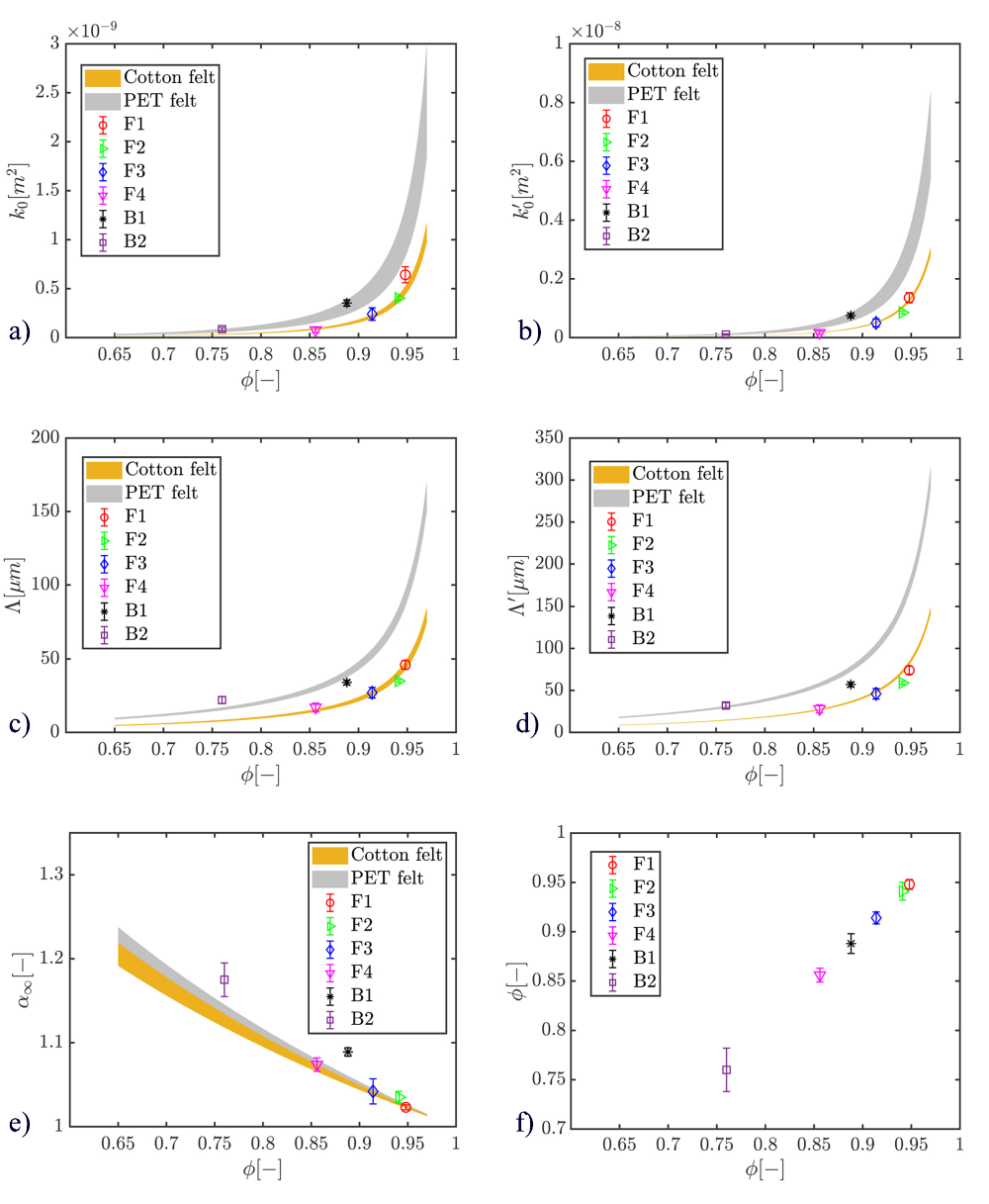}
\caption{\label{fig:transport parameter} Evolution of the transport patameters $k_0$, $k_0'$, $\Lambda$, $\Lambda'$, $\alpha_{\infty}$ with the porosity $\phi$ for three-dimensional random fibrous materials with transversely isotropic structure and a preferred angular orientation $\Omega_{zz}$ depending on the compression rate $n$. Comparison between the predictions of the semi-analytical models Tab. \ref{tab:analytical model} and the data obtained from experiments (symbols). These predictions are obtained using the average microstructurals descriptors in Tab. \ref{tab:Pd_beta} for the cotton felts ($D_v=18.95\pm0.5 \mu m$; $D_{iv}=9.20\pm0.26 \mu m$; $\Omega_{zz}=0.15\pm0.09$; $CV=40.2\pm1.2 \%$) and for the PET felts ($D_v=28.95\pm3.25 \mu m$; $D_{iv}=19.20\pm0.85 \mu m$; $\Omega_{zz}=0.05\pm0.05$; $CV=29.9\pm4.6 \%$) The thick lines correspond to the deviation of either cotton felts (orange) or PET felts (grey).}
\end{figure*}

\begin{figure*}[ht]
\centering
\includegraphics[width=14cm]{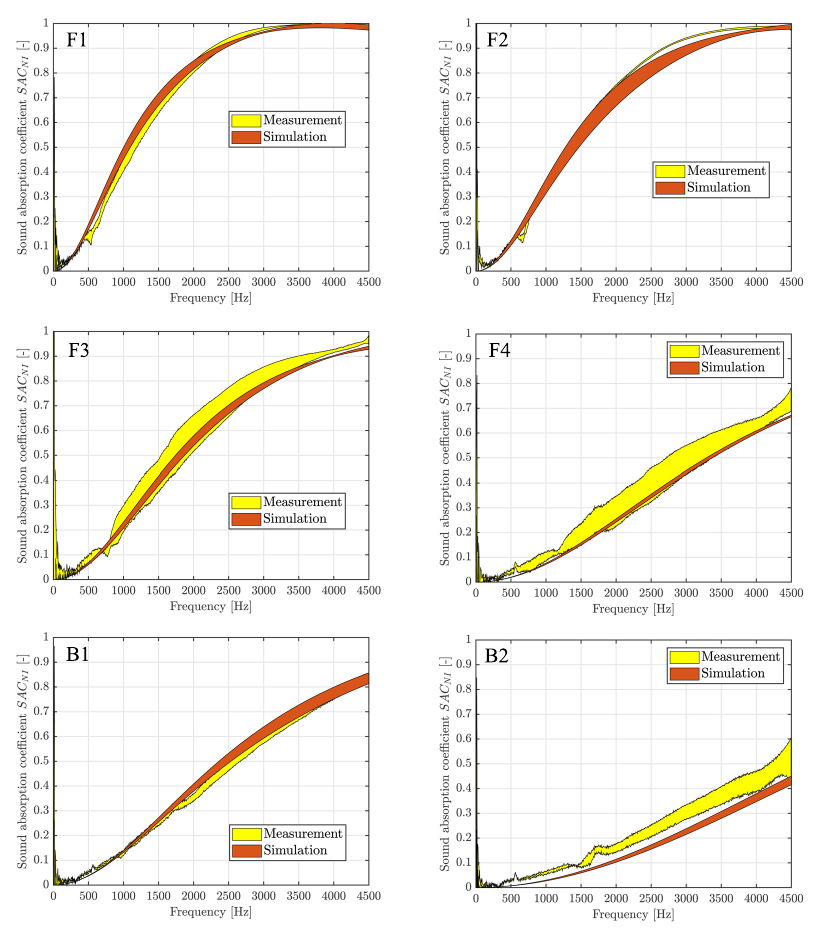}
\caption{\label{fig:Absorption} Comparison between measurements and predictions of the sound absorption coefficient at normal incidence. Sample thickness: F1 - 20.3 $mm$; F2 - 16.1 $mm$; F3 - 11.2 $mm$; F4 - 5.9 $mm$; B1 - 10.3 $mm$; B2 - 4.3 $mm$.}
\end{figure*}

Two different types of comparisons are presented to validate the semi-analytic model in Tab. \ref{tab:analytical model}. The first type of comparisons, shown in Fig. \ref{fig:transport parameter}, concerns the transport properties predicted by the model and their experimental measurements or estimates, presented in Section \ref{sec:results transport properties}. The second type of comparisons, shown in Fig. \ref{fig:Absorption}, concerns the sound absorption coefficient predicted by the model for each felt and its impedance tube measurement obtained from the method presented in Section \ref{sec:method for transport and acoustic}. From these comparisons, several important results can be drawn.  They are listed below.
\begin{itemize}
    \item From the comparisons shown in Fig. \ref{fig:transport parameter}, one can conclude that the proposed semi-analytical model allows nice quantitative predictions of the measured transport properties $k_0$, $k'_0$, $\Lambda$, $\Lambda'$, and $\alpha_{\infty}$. The comparison is good for a wide range of open porosities $(0.760 \leq  \phi \leq 0.948)$ [Tab. \ref{Tab:result}] and for different fiber orientation distributions $(0.01 \leq \Omega_{zz} \leq 0.22)$ [Tab. \ref{tab:Pd_beta}].  It is recalled that fiber orientation is related to the compression ratio that varies in the range $(1 \leq n \leq 3.4)$ for the two families of composite nonwoven fibrous materials (F and B, Tab. \ref{tab:information fiber}).  These two families have a different fiber diameter polydispersity content ($CV \sim 40\%$ for F and $CV \sim 30\%$ for B, Fig. \ref{fig:diameter_angle}b and Tab. \ref{tab:Pd_beta}). Consequently, the overall agreement between the analytical and experimental results supports the validity of the semi-analytical model within, at least, the degrees of fiber diameter polydispersity and orientation studied.  Moreover, this proves that the fiber diameter polydispersity and, to a lesser extent, the orientation of fibers play a leading role in the transport properties of these fibrous composites.
    \item 	Despite a relatively good overall comparison, a few differences are worth discussing.  First, for $k_0'$, we recall here that it was not possible to have a direct measurement of $k_0'$.  Its value is estimated from the identification of $\Lambda'$ (Eq. \ref{eq:k0p estimation}), which is in turn estimated from other measured properties thanks to the Kozeny-Carman formula (Eq.: \ref{eq:Lct estimation}).  Consequently, we must look at the trend of its evolution more than its values.  The same holds for $\Lambda'$ and $\Lambda$.  Second, the predicted value of $k_0$ for F1 departs from the measurement. As explained previously (Section 5.2), the model diverge for high porosity values approaching.
    \item We next explored the sound absorbing behavior at normal incidence in an analytical way using the predicted transport parameters in a JCAL model (\ref{Elementary transport processes and acoustical macro-behavior}) that allowed us to generate the sound absorption coefficient that could be compared directly with experiments (Fig. \ref{fig:Absorption}). This analysis shows that the sound absorption coefficients at normal incidence that are predicted are comparable to those measured experimentally. Together with a close match between the transport parameter values in the experiments and in the models, this and the above results confirm the accuracy of the numerical models and indicate that they capture the essential physics of the viscous fluid-flow, excess temperature, and potential flow velocity field in a polydisperse nonwoven composite and the corresponding transport and sound absorbing properties.
 \end{itemize}
\begin{table*}[ht]
\centering
\begin{tabular}{cccccccc} 

\hline\hline
            & Results & $\phi$            & $\sigma (N.s.m^{-4})$ & $\alpha_\infty$  & $\Lambda (\mu m)$       & $ \Lambda' (\mu m) $ & $k'_0 \times10^{-10} (m^2)$ \\ \hline
\multirow{2}{*}{F1} & Model   & $ 0.948\pm0.005$ & $38358\pm1612 $       & $ 1.022\pm0.002$ & $48\pm5$                & $ 84\pm8$            & $ 12.1\pm1.6$                \\
                   & Exp   & $ 0.948\pm0.005$  & $28684\pm3664 $       & $ 1.023\pm0.003$ & $46\pm3$                & $ 74\pm5$            & $13.6\pm1.7$                  \\ \hline
\multirow{2}{*}{F2} & Model   & $ 0.941\pm0.009$ & $47235\pm3191$        & $ 1.026\pm0.004$ & $42\pm7$ & $ 74\pm12$            & $ 9.9\pm2.1$                \\
                   & Exp   & $ 0.941\pm0.009$  & $45716\pm2553 $       & $ 1.035\pm0.007$ & $35\pm1$                & $ 59\pm2$            & $8.6\pm0.4$                  \\ \hline
\multirow{2}{*}{F3} & Model   & $ 0.914\pm0.006$ & $87776\pm2818$       & $ 1.042\pm0.003$ & $25\pm2$                & $ 47\pm4$            & $ 5.1\pm0.5$                \\
                   & Exp   & $ 0.914\pm0.006$  & $76479\pm20416 $      & $ 1.042\pm0.015$ & $27\pm4$                & $46\pm6$             & $ 5.1\pm1.3$                  \\ \hline
\multirow{2}{*}{F4} & Model   & $ 0.856\pm0.007$ & $242696\pm5784 $      & $ 1.078\pm0.005$ & $15\pm1$                & $ 28\pm2$            & $ 1.6\pm0.1$                \\
                   & Exp   & $ 0.856\pm0.007$  &$235845\pm105324 $    & $ 1.074\pm0.008$ & $17\pm4$                & $ 28\pm6$           & $ 1.7\pm0.8$                  \\ \hline
\multirow{2}{*}{B1} & Model   & $ 0.887\pm0.001$  & $65456\pm2770 $       & $ 1.057\pm0.006$ & $42\pm4$                & $ 79\pm8$            & $ 6.3\pm0.8$                \\
                   & Exp   & $ 0.888\pm0.001$  & $52018\pm4732 $       & $ 1.089\pm0.005$ & $34\pm2$                & $ 57\pm3$            & $ 7.5\pm0.6$                  \\ \hline
\multirow{2}{*}{B2} & Model   & $ 0.764\pm0.022$  & $246602\pm12283 $     & $ 1.144\pm0.021$  & $15\pm2$                & $ 29\pm4$            & $ 1.18\pm0.1$                  \\
                   & Exp   & $ 0.760\pm0.022$   & $213834\pm44998$      & $ 1.175\pm0.02$  & $22\pm2$                & $ 32\pm4$            & $ 1.1\pm0.2$                   \\ \hline\hline
\end{tabular}
\caption{Comparison of semi-analytical (Model) and experimental (Exp) estimates of the transport parameters of cotton and PET felts}
\label{Tab:result}
\end{table*}
\FloatBarrier
\section{Conclusions}
\label{sec:Conclusions}
The objective of this study was to link the macroscale transport and sound absorbing properties of nonwoven fibrous composites with their polydisperse fibrous microstructures and the related visco-thermal dissipation mechanisms. For that purpose, two families of composite nonwovens were manufactured using a thermo-compression process, from either recycled cotton and co-PET fibers or a mix of recycled PET and Co-PET fibers with different classes of fineness, and further compacted with several compression rates. SEM images showed that their random fibrous microstructures exhibited well known transverse isotropy with a preferential orientation of fibers that depended on the compression rate. In addition, regardless of the family of the composite nonwovens, the fibers originating from a recycling process were characterized by a wide distribution of diameters which could be modeled as a Gamma-law, a trend already observed for glass and stone wools. From the fiber scale images of their microstructures, we also saw that the radius of curvature of the fibers was large when compared to the fiber radii, so that the individual fibers could be considered as straight cylinders. The connectivity between two adjacent fibers due to the thermo-compression bounded co-PET process was also visible so as to reasonably assume that fibers could intersect.\\
From these experimental data obtained at fiber scale, fiber network models were proposed to predict the through-plane transport properties of the considered polydieperse nonwoven composites. Two microscale models were established. The first one used volume weighted fiber diameter and the second inverse volume weighting as mean diameters to perform finite element simulations. The results were rationalized in the form of analytical laws that can be easily used for engineering purposes, e. g., to optimize polydisperse fibrous media. The modeling approach emphasised the leading roles of the fiber content, polydispersity and orientation on the macroscale transport and sound absorbing properties of the considered nonwovens. The modeling approach quantitatively well predicted the transport and sound absorbing properties characterized at macro-scale. If the porosity and distributions of fiber diameters and orientations are provided as inputs, we have shown that the predictions of the numerical and analytical models can nicely estimate the transport and sound absorbing properties at normal incidence of random and transversely isotropic polydisperse fibrous media for a large range of porosities and without any adjusted parameter. The identified micro-structural descriptor of the low frequency behavior is in accordance with literature data, i. e., at low polydispersity content, only one fiber diameter is necessary to derive the overall transport parameters characterizing both low and high frequency behaviors, thus suggesting a switch from mono disperse to poly disperse fiber distribution as a new lever to understand and optimize transport and sound absorbing properties. The developed model should be tested accordingly for fiber diameter distributions characterized by very large coefficient of variations.
\section*{Acknowledgments}
We are grateful with several stimulating discussions with two anonymous reviewers. This work was part of a project supported by ANRT and Adler Pelzer Group, Acoustic TechCenter R$\&$D under convention CIFRE No. 2020/0122. The MSME laboratory is part of the LabEx MMCD (Investisements d'Avenir: grant agreement no. ANR-11-MABX-022-01). Partial support for this work was also provided by Université Paris-Est Sup (mobility grant from the ED SIE). The authors also acknowledge the support of the Natural Sciences and Engineering Research Council of Canada (NSERC) [funding with ref. number RGPIN-2018-06113]. We acknowledge Rémy Pires-Brazuna (ICMPE UMR 7189 CNRS)  for SEM imaging of the fibrous samples.
\appendix
\section{Protocol of preparation and cutting of samples prior to the acquisition of SEM images} 
\label{Appendix: Protocol}
For non-conductive materials like cotton and PET felts, a high performance metallizer by cathode sputtering was used, coupled with a magnetron source (Cressington sputter coater 208HR); which made it possible to deposit a conductive film of a few nanometers (controlled by a quartz probe, here a Cressington MTM 20) on the surface of the samples.
To verify the homogeneity of the microstructure, specifically in terms of fiber diameters, two cubic specimens with dimensions of $10$ $mm$ were taken randomly from different locations of the studied panels (provided with dimensions of 210 mm x 297 mm).
On each extracted cubic specimen, SEM images were then acquired to fully scan two horizontal and two vertical planes (situated on opposite faces of the cubic specimens), using a magnification factor of 100 times.
For each plane (four planes of interest on each cubic sample), 10 sub-images were randomly extracted to directly measure the morphological parameters of interest in the fibrous network, using the FiJi software \cite{schindelin2012fiji} with a resolution of $0.56$ $\mu m$ per pixel.
These parameters include the diameters of the fibers and their orientation angles in the horizontal and vertical planes, as shown in Fig. \ref{fig:SEM}.
\section{ Evaluation of the bias induced from the projection process}
\label{Appendix: Calculation of the bias induced from the projection process}
\begin{figure*}[ht]
\centering
\includegraphics[width=9cm]{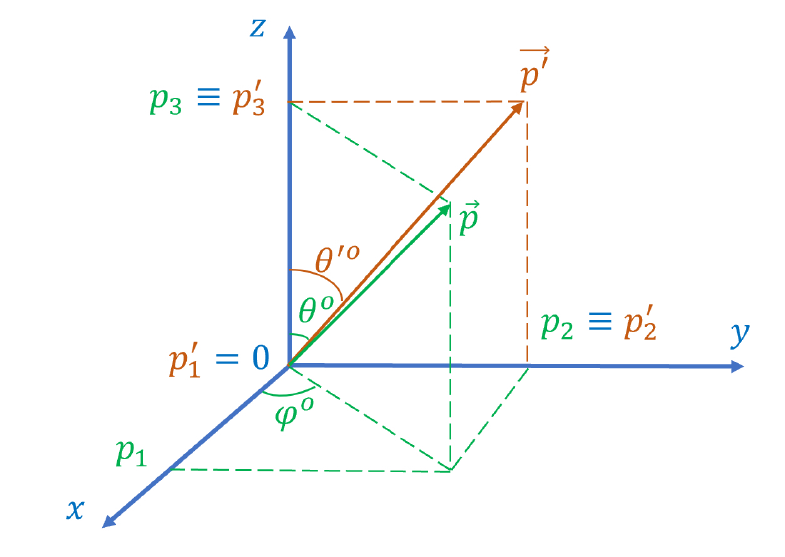}
\caption{Three-dimensional representation of a vector p characterizing the angular orientation of a fiber ($\varphi$, $\theta$) and its projection over orthogonal plans $(Ox, Oy)$ and $(Oy, Oz)$ with, as a result, apparent angular orientation ($\varphi$, $\theta'$).}
\label{fig15} 
\end{figure*}

\begin{figure*}[ht]
\centering
\includegraphics[width=14cm]{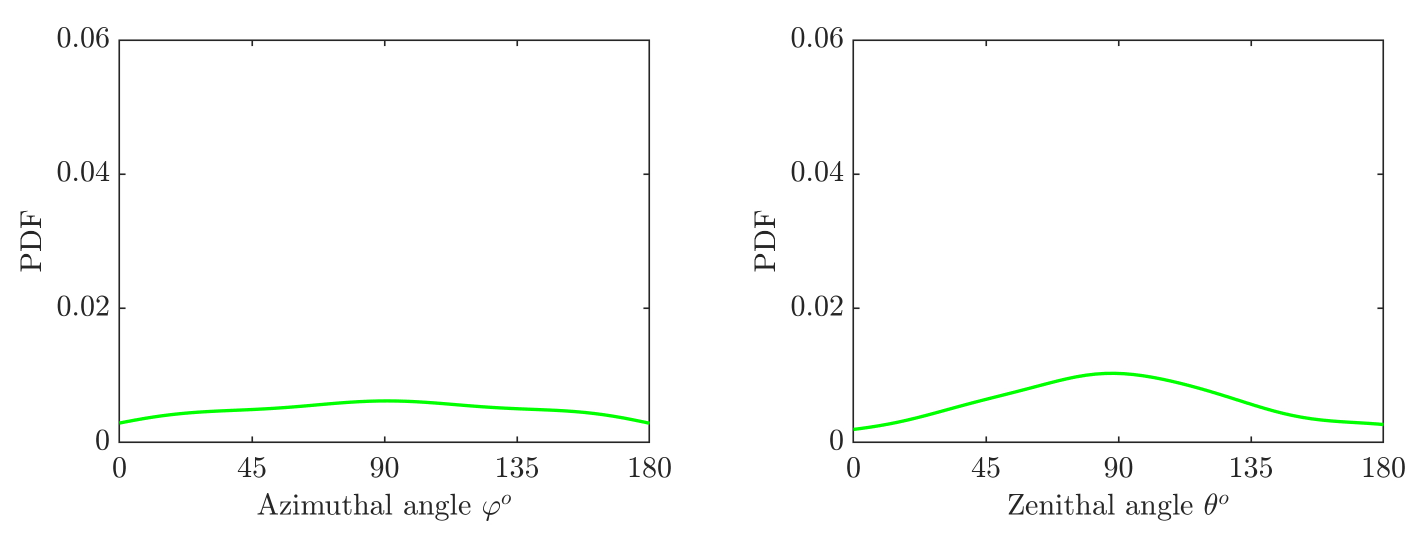}
\caption{Angular distributions $\varphi$ and $\theta$ of a reference configuration generated in a three-dimensional space before any projection process.}
\label{fig16} 
\end{figure*}

\begin{figure*}[ht]
\centering
\includegraphics[width=14cm]{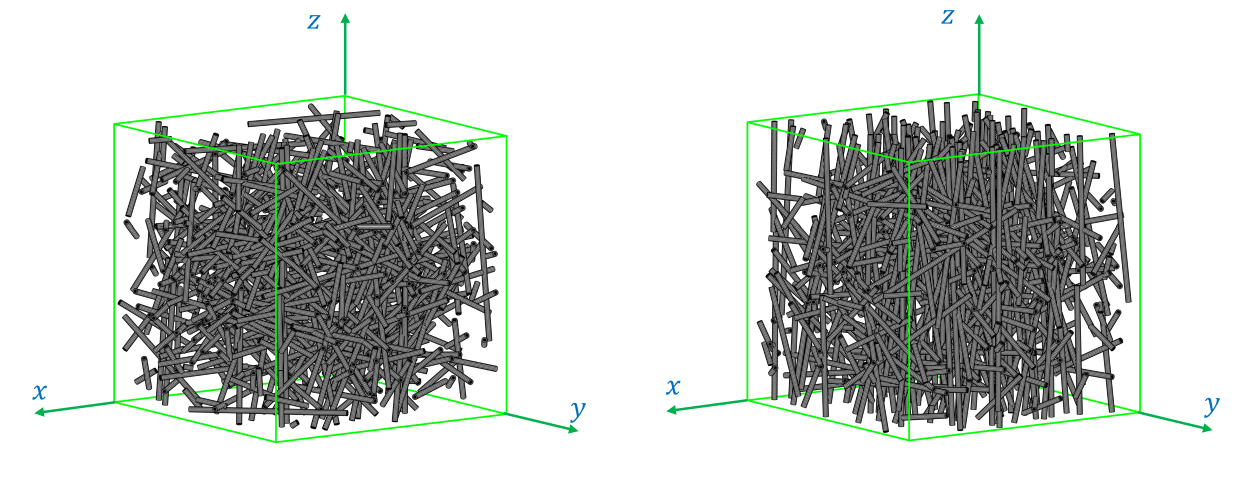}
\caption{Comparison between an isotropic fibrous structure (left panel) with $\Omega_{zz}=0.34$ and the corresponding microstructure obtained after characterization (right panel) using two-dimensional orthogonal images [$(Ox, Oy)$ and $(Oy, Oz)$] with  $\Omega_{zz}=0.49$.}
\label{fig17} 
\end{figure*}

\begin{figure*}[ht]
\centering
\includegraphics[width=9cm]{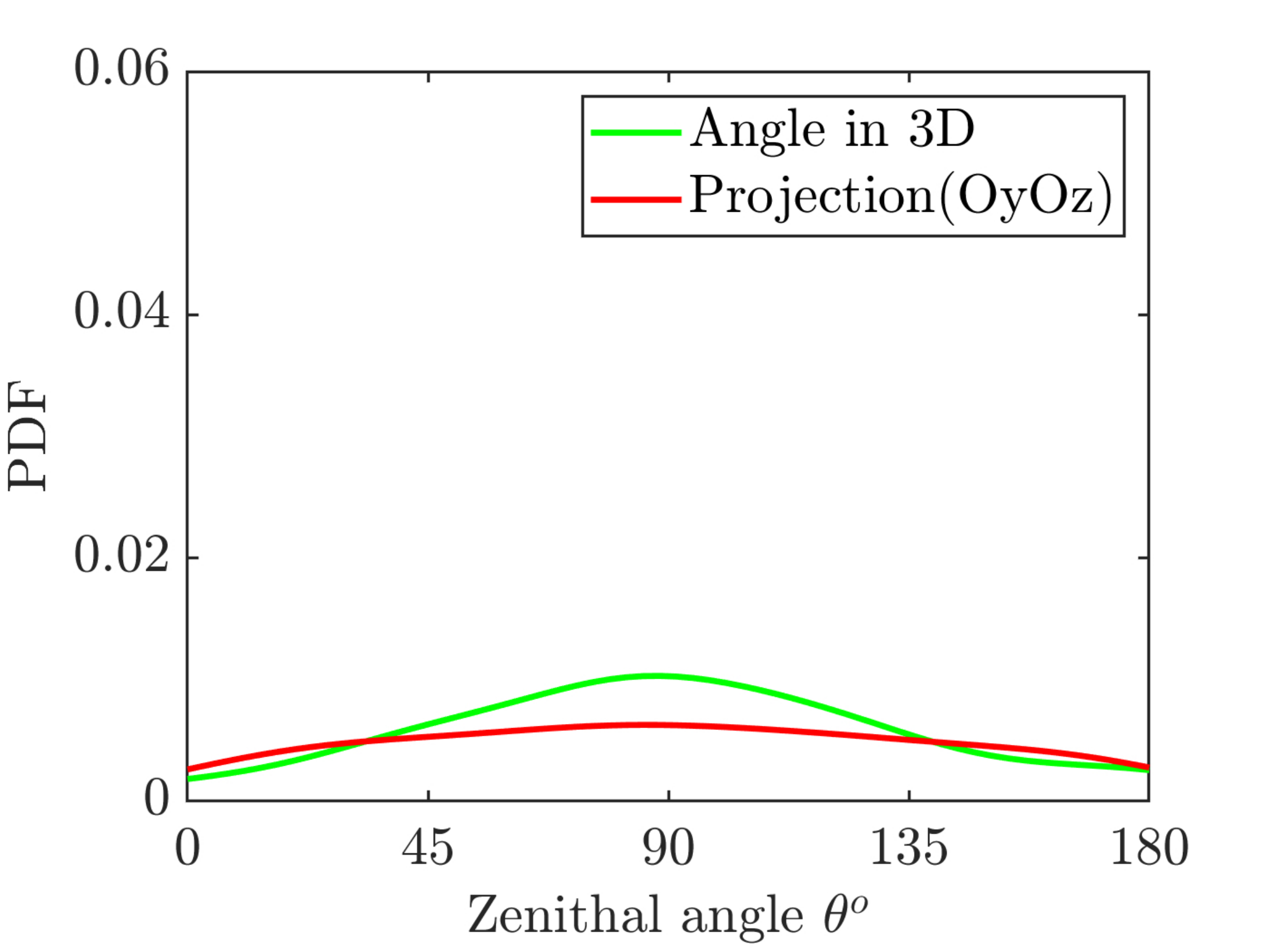}
\caption{Comparison between the angular distribution $\theta$ of an isotropic fibrous microstructure and the approximate distribution $\theta'$ obtained after characterization using two-dimensional orthogonal images [$(Ox, Oy)$ and $(Oy, Oz)$].}
\label{fig18} 
\end{figure*}

Let us analyze the differences between the orientation of a fiber given by its vector p in a three-dimensional space and the corresponding projected information collected on projected plans $[(Ox, Oy), (Oy, Oz)]$.
Obviously, there is no bias on the determination of $\varphi$ from the projection of $\overrightarrow{p}$ on $(Ox, Oy)$, since $\varphi$ is already defined on this plan.
The reverse is not true regarding the characterization of $\theta$ from two-dimensional images : there is a difference between $\theta$ defined by the angle between $Oz$ and $\overrightarrow{p}$; and $\theta'$ defined by the angle between $Oz$ and the projection of $\overrightarrow{p}$ over $(Oy, Oz)$, say $\overrightarrow{p'}$ (Fig. \ref{fig15}).
To gain insight into the bias introduced by our characterization process, we need to further study the differences statistically obtained through this projection procedure.
Recall that the vector $\overrightarrow{p}$ in defined in a three-dimensional coordinate system as
\begin{align}
\label{eq:B1}
    \overrightarrow{p} = \begin{pmatrix} p_1 \\ p_2 \\ p_3 \end{pmatrix} = \begin{pmatrix} p \sin \theta \cos \varphi \\ p \sin \theta \sin \varphi \\ p \cos\theta \end{pmatrix}.
\end{align}
The projection $\overrightarrow{p'}$ of the vector  $\overrightarrow{p}$  on $OyOz$ has the following coordinates
\begin{align}
\label{eq:B2}
    \overrightarrow{p'} = \begin{pmatrix} p'_1 \\ p'_2 \\ p'_3 \end{pmatrix} = \begin{pmatrix} 0 \\ p' \sin \theta' \\ p' \cos \theta' \end{pmatrix},
\end{align}
where we identified the equality, $p_3=p'_3$ (Fig. \ref{fig15}), leading to the following equation:
\begin{align}
    p \cos \theta=p' \cos \theta',     \label{eq:B3}\\
    p' = p \frac{\cos \theta}{\cos \theta'}. \label{eq:B4}
\end{align}
Also noticeable is the following equality,  $p_2=p'_2$ (Fig. \ref{fig15}), from which we get using Eqs. \ref{eq:B1} and \ref{eq:B2}
\begin{align}
    \label{eq:B5}
    p \sin \theta \sin \varphi = p' \sin \theta'.
\end{align}
By using Eq. \ref{eq:B4}, it follows that
\begin{align}
    p \sin\theta \sin\varphi = p \frac{\cos\theta}{\cos\theta'} \sin\theta'   \label{eq:B6}\\
    \Leftrightarrow p \frac{\sin\theta}{\cos\theta} \sin\varphi = p \frac{\sin\theta'}{\cos\theta'} \label{eq:B7}\\
     \Leftrightarrow \tan\theta\sin\varphi=\tan\theta' \label{eq:B8}
\end{align}
We therefore obtained the relationship sought between $\theta'$ and $\theta$ quantifying the bias introduced by the projection process over $OyOz$ :
\begin{align}
    \label{eq:B9}
    \theta'=\tan^{-1}(\tan\theta\sin\varphi).
\end{align}
Let us now quantify in a systematic manner the differences obtained between $\theta$ and $\theta'$ for a three-dimensional fibrous network characterized by an isotropic structure. The corresponding angular distributions ($\varphi$, $\theta$) are provided throughout Fig. \ref{fig16}, together with an illustration of the geometry of the fibrous microstructure throughout Fig. \ref{fig17}.\\
Quantitatively, we recall that the differences between of $\theta$ and $\theta'$ emerge from the projection process and can be determined from Eq. \ref{eq:B9}.  In Fig. \ref{fig18}, we display and compare the original angular distribution $\theta$ and the one corresponding to $\theta'$ obtained after projection of $\overrightarrow{p}$ over the $OyOz$ plan.
From Figs. \ref{fig17}-\ref{fig18} and the afore mentioned calculations, we see that the projection process tends to slightly overestimate the determination of the zenithal angle in the angular ranges $[0^o, 32.5^o]$ and $[141.2^o, 180^o]$; while the zenithal angle might be underestimated in the remaining angular domain $[32.5^o, 141.2^o]$ with a maximum statistical deviation typically observed for horizontal fibers lying in the $OxOy$ plan. Both of these tendencies are reflected by a projection bias which tends to artificially verticalize the reconstructed microstructures (horizontal fibers are slightly underrepresented while there is a small overrepresentation of vertical fibers). A procedure is under development to avoid such a bias, and will be the purpose of a future communication. Let us mention however that the angular orientation of fibers seems relatively well captured in general.\\
\section{Experimental approach used to estimate the viscous and thermal characteristic lengths}
\label{Appendix: Estimation characteristic lengths}
The so-called Kozeny-Carman resistivity formula, introduced by Henry \textit{et al.} \cite{Henry1995} in their Eq. (15), is given by:
\begin{align}
    \sigma_{KC} = \frac{8\alpha_{\infty}\eta}{\phi\Lambda_{est}^{'2}}, \label{eq:sigma_KC}
\end{align}
where $\Lambda'_{est}$ is a characteristic dimension. Typically, we can assume that the value of $\Lambda'_{est}$ is between $\Lambda$ and $\Lambda'$, and that $\sigma_{KC}$ is an estimate of $\sigma$. From the Kozeny-Carman formula, a value of $\Lambda'_{est}$ could be obtained using experimental measurements of $\phi$, $\sigma$, and $\alpha_{\infty}$. Therefore, $\Lambda'_{est}$ corresponds to the following equation:
\begin{align}
   \Lambda'_{est}= \sqrt{\frac{8\alpha_{\infty}\eta}{\phi\sigma}}. \label{eq:Lct estimation}
\end{align}
For a typical porous material, assuming macroscopic homogeneity, the following inequality $\Lambda\leq\Lambda'_{est}\leq\Lambda'$ is expected. As a first approximation, the simulated ratio $r=\Lambda'/\Lambda$ can be used to deduce $\Lambda'_{est}$ from $\Lambda_{est}$. The following formula is applied:
\begin{align}
   \Lambda_{est}= \frac{\Lambda'_{est}}{r}, \label{eq:Lcv estimation}
\end{align}
Here, we used $r=$ 1.61, 1.69, 1.70, 1.65, 1.68 and 1.45 corresponding to the simulated values for F1, F2, F3, F4, B1, and B2, respectively. 
\section{ Smallest channels in polydisperse fibrous structures}
\label{Appendix: Smallest channels in polydisperse fibrous structures}
As the frequency increases to the point where the viscous skin depth, $\delta_v$, becomes small compared to the pore dimensions, the microscopic fluid-flow pattern crosses over to potential-flow except within a boundary layer of thickness $\delta_v$ at the pore walls and it has been shown that [Johnson \textit{et al.} \cite{johnson1986new}; Johnson \textit{et al.} \cite{johnson1987theory}]
\begin{align}
\label{eq:D1}
\lim_{{\omega \to \infty}} \tilde{k}(\omega) = \frac{i \eta \phi}{\alpha_{\infty} \rho_0 \omega} \left[1 - \left(\frac{i \eta}{\rho_0 \omega}\right)^{1/2} \frac{2}{\Lambda}\right].
\end{align}
We emphasize that this result, which determines the high-frequency acoustics behavior of fluid-saturated solid microstructures, involves parameters $\alpha_{\infty}$ and $\Lambda$ which derive from the solution of the microscopic potential-flow equations \ref{eq:electric}.\\
In practice, $\alpha_{\infty} \approx 1$ for fibrous media, a simple array of thin solid fibers ($\phi \to 1$) does not lead to a significant dispersion of the microscopic potential-flow velocities [$\langle \textbf{E} \cdot \textbf{E} \rangle_f \geq \langle \textbf{E} \rangle_f \cdot \langle \textbf{E} \rangle_f$, but $\langle \textbf{E} \cdot \textbf{E} \rangle_f \approx \langle \textbf{E} \rangle_f \cdot \langle \textbf{E} \rangle_f$; in Eq. \ref{eq:tor}].\\
$\Lambda$ is essentially the pore-to-surface volume ratio of the pore-solid interface in which each area or volume element is weighted according to the local value of the microscopic potential-flow velocity $\textbf{E}$. The volume flow rate having an imposed value, $\Lambda$ is obviously dominated by the smallest channels where the microscopic potential-flow velocities reach the larger values.\\
Therefore, an experimental measure of the high-frequency acoustic behavior of a locally heterogeneous fibrous medium is expected to be dominated by the $\Lambda$ parameter which is very sensitive to the smallest channels of the polydisperse microstructure.\\
One may expect the local characteristic size of these smallest channels to be captured, at least approximately, by the reconstruction of a REV which promotes the smallest fibers of the network [Eq. \ref{eq:Div}] at known and given porosity $\phi$.\\
\section{Geometrical reconstruction}
\label{Appendix:Geometrical reconstruction}
\begin{figure}[ht]
\centering
\includegraphics[width=9cm]{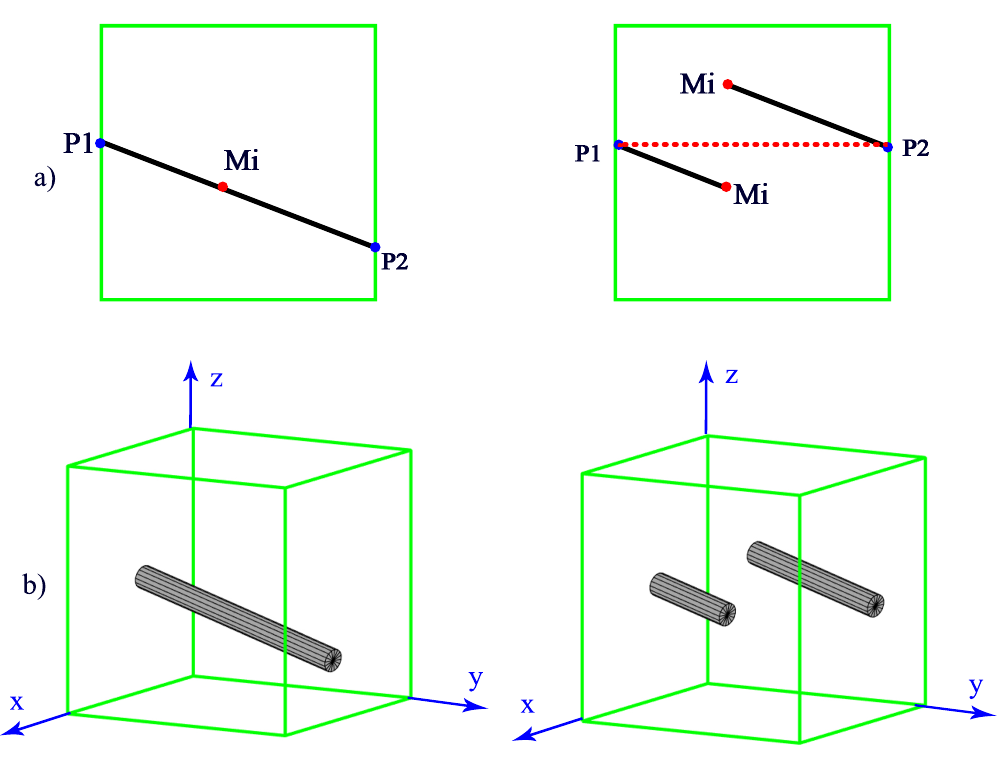}
\caption{\label{fig:periodic} Illustration of some important steps by which a representative volume element with periodic boundaries can be constructed.}
\end{figure}

\begin{figure*}[ht]
\centering
\includegraphics[width=14cm]{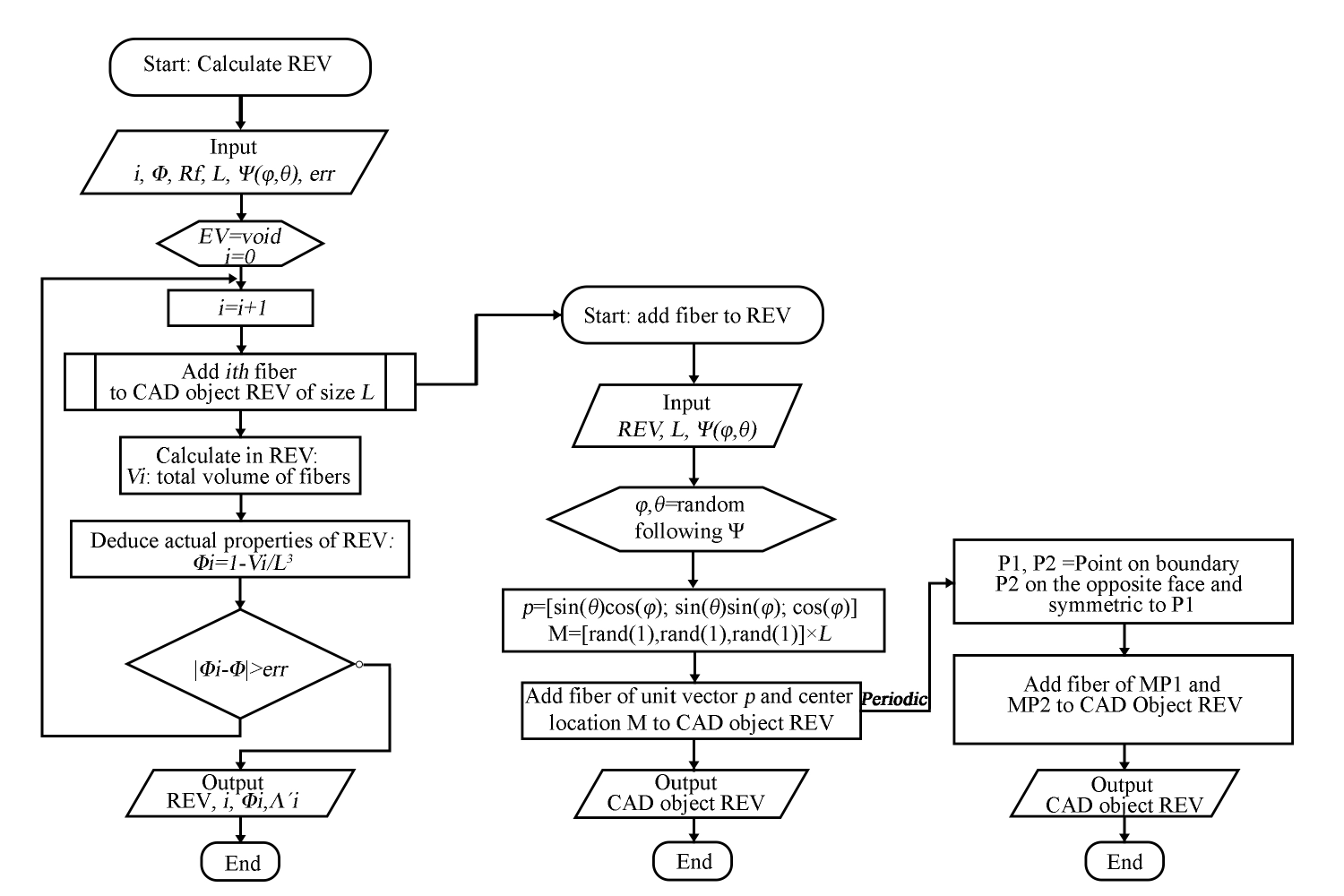}
\caption{\label{fig:algo} Algorithm used to calculate the domain size in order to reconstruct microstructures of the random fibrous materials under study with periodic boundary conditions.}
\end{figure*}

Based on the results of microstructure characterization, a random fibrous network is reconstructed as follows :
\begin{enumerate}
\item   A random point is chosen in a unit cube of known size $L$ (the unit cell).
\item  From this random  point $M_i$, a vector $\overrightarrow{p}$ is determined, which passes through this random point (having as zenithal $\theta$ and azimuthal $\varphi$ angles, randomly selected values from the measured probability density functions).
\item Based on the knowledge of $\overrightarrow{p}$, the  coordinate of the intersecting points $P_1P_2$ with the unit cube are derived.
\item Next, the segment $P_1 P_2$ is cut at $M_i$, from which one can obtain continuity of the solid phase  on the opposite faces of the unit cube. This is done by translation of a sub-segment $M_i P_2$.  For instance, Fig. \ref{fig:periodic}a illustrates this procedure. Here, $M_i P_2$ is translated to  ensure continuity of $P_1$ and $P_2$ (by horizontal  translation of the unit cube).
\item Knowing the fiber diameter distribution obtained from measurements, a fiber radius is then randomly drawn from the corresponding Gamma fit distribution (Fig. \ref{fig:periodic}b).
\end{enumerate}
The algorithm which is reported in Fig. \ref{fig:algo} allows iterative alteration of the fiber number $N_f$ and the domain size $L_i$ until porosity is converged towards the experimentally determined value. By applying the algorithm with 100 iterations for each domain size $L/D_m$, the result displayed in Fig. \ref{fig:domainsize} shows that it is possible to control both the average porosity and the standard deviation of a reconstructued three-dimensional fibrous structure. $L_i$ was chosen to ensure that the ratio $\epsilon$ of the standard deviation over the mean value of the targeted porosity is less than $0.1\%$. 
\FloatBarrier
\section{Elementary transport processes and acoustical macro-behavior}
\label{Elementary transport processes and acoustical macro-behavior}
\subsection*{Elementary transport processes}
\label{subAppendix: Elementary transport processes}
In this section, we focus on identifying macroscopic transport properties by addressing local equations with adequate boundary conditions. These equations are classically derived from an asymptotic analysis.\\
Note that, the open porosity $\phi$ and the thermal characteristic length $\Lambda'$ are purely geometric parameters that can be directly calculated from the microstructure and determined by integration: 
\begin{align}
    \phi=\frac{\int_{\Omega_f}{dV}}{\int_{\Omega}{dV}}, \label{eq:phi}\\
    \Lambda'=2\frac{\int_{\Omega_f}{dV}}{\int_{\partial\Omega}{dS}}, \label{eq:TCL}
\end{align}
where $\Omega$ is the volume element (VE), $\Omega_f$ is the fluid volume and $\partial\Omega$ denotes the solid border of a solid element.\\
The remaining transport parameters are determined numerically by applying spatial averaging to the solution fields corresponding to the problems mentioned below.
\begin{enumerate}
    \item Viscous permeability\\
At low frequencies, also known as the static regime, viscous forces are dominant. The low Reynolds number flow of an incompressible Newtonian fluid in this regime is governed by the steady-state Stokes equation:
\begin{align}
\begin{split}
   \eta\Delta\textbf{v}-\nabla p=-\nabla p^m \hspace{0.5cm} \text{in} \hspace{0.5cm} \Omega_{f} ,\\ 
    \nabla\cdot\textbf{v}=0 \hspace{0.5cm} \text{in} \hspace{0.5cm} \Omega_{f} ,\\
    \textbf{v}=0 \hspace{0.5cm} \text{on} \hspace{0.5cm} \partial\Omega,\\
    \textbf{v}\quad\text{and} \, p \,\text{are}\, \Omega-\text{periodic}; \label{eq:stokes}
    \end{split}
\end{align}
where $\textbf{v},p$, and $\eta$ are the velocity, pressure, and viscosity of the fluid, respectively. The term $\nabla p^m$ is a macroscopic pressure gradient acting as a driving force, $\partial\Omega$ is the fluid-solid interface.
The macroscopic pressure gradient is specified in the form,
\begin{align}
\nabla p^m=|\nabla p^m|\textbf{e}. \label{eq:pm}
\end{align}
Since the Eq. \ref{eq:stokes} is linear, it can be shown that
\begin{align}
\phi\left<\textbf{v}\right>=-\frac{\textbf{K}}{\eta}\cdot\nabla p^m, \label{eq:phi_v}
\end{align}
where $\textbf{K}$ is a positive-definite symmetric tensor, the symbol $\left<\bullet\right>$ indicates a fluid-phase averaging, that is 
\begin{align*}
\left<\bullet\right>=\frac{1}{\Omega_f}\int_{\Omega_f}{\bullet \, dV}.
\end{align*}
The static permeability $k_0$ along the direction specified by the unit vector is calculated as,
\begin{align}
k_0=(\textbf{K}\cdot\textbf{e})\cdot\textbf{e}=-\frac{\eta\phi}{|\nabla p^m|}\left<\textbf{v}\right>\cdot\textbf{e}. \label{eq:k_0}
\end{align}
\item Viscous characteristic length and tortuosity\\
At high frequencies, when $\omega$ is large enough, inertial forces dominate over viscous forces. The fluid tends to behave as an ideal fluid, having no viscosity. In this case, the inertial flow problem is analogue to the problem of electric conduction of a conducting fluid saturating an insulating porous structure :
\begin{align}
\begin{split}
    \textbf{E}=\textbf{e}-\nabla\varphi \quad \text{in}\quad\Omega_f ,\\
    \nabla\cdot\textbf{E}=0 \quad \text{in}\quad\Omega_f ,\\
    \textbf{E}\cdot\textbf{n}=0 \quad \text{on}\quad\partial\Omega,\\
    \varphi\:\text{is}\:\Omega-\text{periodic}, \label{eq:electric}
    \end{split}
\end{align}
where  $\textbf{e}$ is a global unit electric field, while $\textbf{E}$ is the electric field solution of the boundary problem, $-\nabla\varphi$ is the scalar electrostatic potential and \textbf{n} is local unit normal vector directed into the pore space.\\
Then, the components  of the high frequency tortuosity tensor $\alpha_{\infty ij}$ can be obtained from\\
\begin{align}
    e_i= \alpha_{\infty ij}\left<E_j\right>. \label{eq:tor_ij}
\end{align}
In the case of isotropy, the components of the tensor simplify to the diagonal form $\alpha_{\infty ij}=\alpha_{\infty}\delta_{ij}$. The tortuosity can also be determined by calculating the mean square value of the local electric field through
\begin{align}
    \alpha_\infty=\frac{\left<\textbf{E}\cdot\textbf{E}\right>}{\left<\textbf{E}\right>\cdot\left<\textbf{E}\right>}.\label{eq:tor}
\end{align}
The viscous characteristic length $\Lambda$ can also be determined (for an isotropic medium) by 
\begin{align}
     \Lambda=2\frac{\int_{\Omega_f}{\textbf{E}\cdot\textbf{E} \, dV}}{\int_{\partial\Omega}{\textbf{E}\cdot\textbf{E} \, dS}}. \label{eq:VCL}
\end{align}
\item Thermal permeability\\
Under the excitation of an external, harmonic source, with perfect absorbing conditions at the fluid-solid interface, the static thermal permeability is obtained from the equation
\begin{align}
   k'_0=\phi\left<u\right>, \label{eq:k0p}
\end{align}
where the scaled, $\Omega$-periodic temperature field $u$, is the solution to the Poisson equation
\begin{align}
\begin{split}
  \Delta u=-1 \quad \text{in}\quad\Omega_f,\\
  u=0 \quad \text{on}\quad\partial\Omega. \label{eq:up}
  \end{split}
\end{align}
Here $u$ is presumed to be periodic with a period $L_i$ across the three spatial directions. The parameter $k'_0$ is a positive definite scalar that is solely dependent on the  geometry of the medium.
\end{enumerate}
\subsection*{Acoustical macro-behavior}
\label{subAppendix: Acoustic macroscopic behavior}
Significant semi-phenomenological models with visco-thermal dissipation mechanisms were  developed by Johnson \textit{et al.} \cite{johnson1987theory} and Lafarge \textit{et al.} \cite{lafarge1997dynamic}. In these works, the assumption of a rigid solid skeleton was made $a priori$. Johnson \textit{et al.} and Lafarge \textit{et al.} proposed that two general expressions for the frequency-dependence of the visco-inertial and thermal exchanges between the frame and the saturating fluid can be established with two sets of parameters ($\Lambda$, $k_0$, $\alpha_{\infty}$, $\phi$) and ($\Lambda'$, $k_0'$, $\phi$). The model is consistent with the frequency dependence of the first two leading terms of the exact result for high frequencies, but only one term for low frequencies. Both numerical simulations and experiments have demonstrated that the model by Johnson \textit{et al.} and Lafarge \textit{et al.}, known as the JCAL model, is very robust (although not exact). In this section, we provide a summary of the JCAL model, as a mean of prediction of the sound absorption of polydisperse fibrous media.\\
For porous materials having a rigid and motionless skeleton, the equivalent dynamic mass density $\tilde\rho_{eq}(\omega)$ and the equivalent dynamic bulk modulus $\tilde K_{eq}(\omega)$ of the material are computed as
\begin{align}
    \label{eq:rho_w}
   \tilde\rho(\omega)=\frac{\alpha_{\infty}\rho_{0}}{\phi}\left[1+\frac{\phi\sigma}{i\omega\alpha_{\infty}\rho_{0}}\sqrt{1+i\frac{4\alpha^{2}_{\infty}\eta\rho_{0}\omega}{\sigma^{2}\Lambda^{2}\phi^{2}}}\right],
\end{align}
and\\
\begin{align}
\begin{split}
    \label{eq:K_w}
   \tilde K_{eq}(\omega)=\\
  & \frac{\gamma P_0/\phi}{\gamma-(\gamma-1)\left[1-i\frac{\phi\kappa}{k'_{0}C_{p}\rho_{0}\omega}\sqrt{1+i\frac{4k_{0}^{'2}C_{p}\rho_{0}\omega}{\kappa\Lambda'^{2}\phi^{2}}}\right]^{-1}}.
   \end{split}
\end{align}
In these equations, $\sigma=\mu/k_{0}$ is the (through plane) airflow resistivity, $\rho_{0}$ is the density of air, $P_0$ the atmospheric pressure, $\gamma =C_p/C_v$ the ratio of heat capacities at constant pressure and volume, $i$ the imaginary unit and $\omega=2\pi f$ the angular frequency. The wave number  $\tilde k_{eq}(\omega)$ and the characteristic impedance $\tilde Z_{eq} (\omega)$ are then given by:
\begin{align}
    \tilde k_{eq}(\omega)=\omega \sqrt{\tilde\rho_{eq}(\omega)/\tilde K_{eq}(\omega)}, \label{eq:kw}\\
    \tilde Z_{eq}(\omega)= \sqrt{\tilde\rho_{eq}(\omega)\tilde K_{eq}(\omega)}. \label{eq:Zw}
\end{align}
The normal incidence surface impedance is expressed by
 \begin{align}
   \tilde Z_s=-i\tilde Z_{eq} \cot{(\tilde k_{eq}L_s)}. \label{eq:Zs_NI}
\end{align}
The sound absorption coefficient at normal incidence of thickness $L_s$ follows:
 \begin{align}
    SAC_{NI}=1-\bigg |\frac{\tilde Z_s-Z_0}{\tilde Z_s+Z_0}\bigg |^2, \label{eq:SAC_NI}
\end{align}
where $Z_0 = \rho_0c_0$ is the impedance of the air, $c_0$ is the sound speed in air.
\section{ Characteristic transition frequencies}
\label{Appendix: Characteristic transition frequencies}
The viscous transition frequency $f_v$ characterises the transition between the high and low frequency limits of the dynamic viscous permeability $k_{(\omega)}$ (Johnson \textit{et al.} \cite{johnson1986new}). One could estimate $f_v$ using the following simple formula
 \begin{align}
     f_v=\frac{\phi\sigma}{2\pi\rho_0\alpha_{\infty}}, \label{eq:Bv}
 \end{align}
where $\rho_0=1.213\times 10^3 (kg/m^3)$ is the density of air at rest and normal conditions. Here, low frequency means $f\ll f_v$, whereas high frequency corresponds to  $f\gg f_v$.\\
 As a thermal counterpart of $f_v$, thermal transition frequency $f_t$ characterises the transition between high and low frequency limits of the dynamic thermal permeability $k'(\omega)$ (Lafarge \textit{et al.} \cite{lafarge1997dynamic}):
 \begin{align}
     f_t=\frac{\phi\kappa}{2\pi k'_0 C_p}, \label{eq:Bt}
 \end{align}
where $\kappa=2.5\times 10^{-2} (W/m\cdot K)$ is the air heat conductivity, $ C_p=1.219\times 10^3 (J/K)$ is the Isobaric heat capacity of air.\\
Low frequencies refers to $f\ll f_t$ whereas high frequencies must be understood as $f\gg f_t$.\\
From  these simple equations [Eqs. (\ref{eq:Bt})-(\ref{eq:Bv})] and the results in Tab. \ref{Tab:result}, the characteristic transition frequencies of the materials studied are calculated and reported throughout Tab. \ref{tab:transition frequency}. These values are useful to show that the high frequency behavior is barely measurable with a standard impedance tube.
\begin{table}[ht]
\centering
\begin{tabular}{ccc}
\hline\hline       
Samples  & $f_v (Hz)$       & $f_t(Hz)$        \\ \hline
F1       & $3488\pm446$         & $4761\pm610$       \\
F2       & $5454\pm312$         & $7491\pm439$       \\
F3       & $8802\pm2353$         & $12431\pm333$      \\
F4       & $24664\pm11018$        & $34925\pm11908$     \\
B1       & $5565\pm510$         & $8051\pm750$       \\
B2       & $18147\pm3867$        & $49614\pm10766$      \\ \hline\hline
\end{tabular}
\caption{Estimation of the characteristic transition frequencies of cotton felts and PET felts.}
 \label{tab:transition frequency}
\end{table}
\FloatBarrier
\bibliographystyle{elsarticle-num-names} 
\bibliography{References}

\end{document}